\documentclass[referee]{aastex63}

\usepackage{amsmath}
\usepackage{amssymb}
\usepackage{latexsym}
\usepackage{graphicx}
\usepackage{natbib}
\usepackage{color}
\bibpunct{(}{)}{;}{a}{}{,}

\usepackage{amstext}

\begin{document}

\title{Alfv\'enic waves in the inhomogeneous solar atmosphere}




\author{R. J. Morton}\email{richard.morton@northumbria.ac.uk}
\author{R. Sharma}
\author{E. Tajfirouze}
\author{H. Miriyala}
\affiliation{Department of Maths, Physics and Electrical Engineering, Northumbria University, UK}



\begin{abstract}
The solar atmosphere is known to be replete with magneto-hydrodynamic wave modes, and there has been significant investment in understanding how these waves propagate through the Sun's atmopshere and deposit their energy into the plasma. The waves' journey is made interesting by the vertical variation in plasma quantities that define the solar atmosphere. In addition to this large-scale inhomogeneity, a wealth of fine-scale structure through the chromosphere and corona has been brought to light by high-resolution observations over the last couple of decades. This fine-scale sturcture represents inhomogeneity that is thought to be perpendicular to the local magnetic fields. The implications of this form of inhomogeneity on wave propagation is still being uncovered, but is known to fundamentally change the nature of MHD wave modes. It also enables interesting physics to arise including resonances, turbulence and instabilities. Here we review some of the key insights into how the inhomogeneity influences Alfv\'enic wave propagation through the Sun's atmosphere, discussing both inhomogeneities parallel and perpendicular to the magnetic field.
\end{abstract}







\section{Introduction}\label{sec1}
Ever since it was established that the coronal plasma had temperatures in excess of a million degrees 
\citep[based on the emission lines from forbidden transitions of highly ionized atoms,][]{Grotian_1939,edlen1941,alfven_1941,1943ZA.....22...30E}, heating of the Sun’s atmosphere or corona has remained an intriguing problem in solar and stellar astronomy. As one of the promising answers to this dilemma, Alfv\'enic waves have long attracted the attention of the community as a mechanism that could not only answer the coronal heating but can also explain the acceleration of solar wind \citep[e.g.,][]{Arregui_2015, Van_Doorsselaere_2020b}. A particular reason for the interest in this wave mode, as compared to the other magnetohydrodynamic (MHD) waves, is it's highly incompressible nature that makes it difficult for the Alfv\'enic waves to dissipate their energy. This property enables these waves to transfer the mechanical energy from the photosphere, where they are generated via the interaction of the magnetic fields with the convective motions, up into the corona and out into the heliosphere. However, the journey of the Alfv\'enic waves through the Sun’s atmosphere is not straightforward, with the inhomogeneous nature of the plasma leading to many interesting wave-related phenomena. Here we do not just refer to inhomogeneities along the magnetic field, but also the fine-scale perpendicular structuring that is observed throughout the chromosphere and corona.

\begin{figure}[!t]
\centering
\includegraphics[scale=0.1]{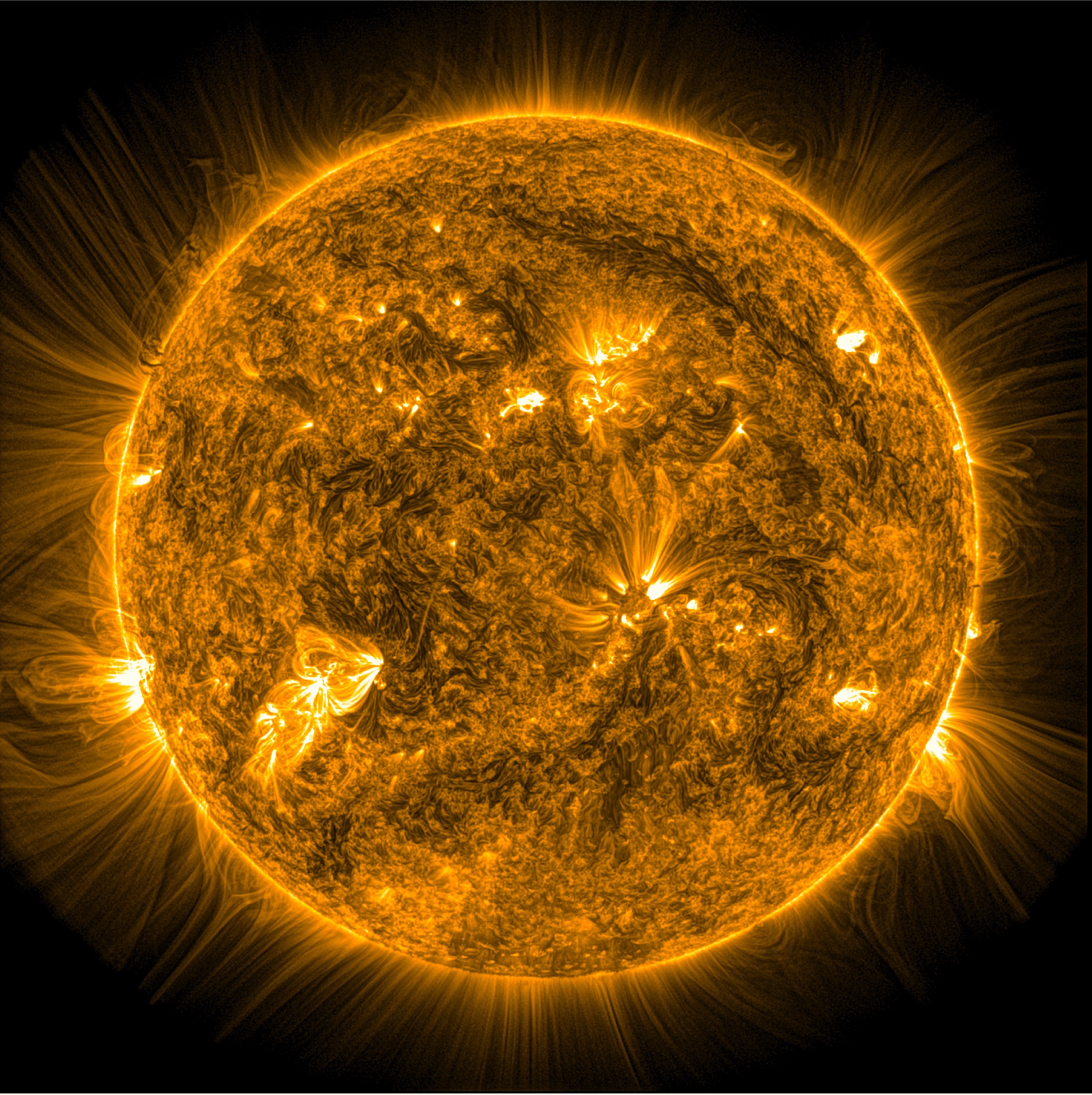}
\caption{An EUV image of the Sun's corona taken with the Solar Dynamic Observatory. The image reveals the fine-scale striations throughout the upper layer of the Sun's atmosphere, indicating the highly inhomogeneous nature of the coronal plasma. The dynamic range of the image has been altered to enable the faint plasma emission around the limb to be visible. }\label{fig:corona}
\end{figure}

An example of such perpendicular structuring is visible in the Extreme Ultra-Violet (EUV) image of the corona from NASA's Solar Dynamic Observatory (SDO) displayed in Figure~\ref{fig:corona}. Present in the image are: a myriad of loop-like structures that emanate from the Active Regions; {comparatively fainter loops in the quiet Sun (best seen off-limb) that are the base of larger structures extending into the heliosphere (e.g., streamers, pseudo-streamers)}; and near-radially oriented striations extending out of the field of view in the quiet Sun and polar regions (again best seen off-limb). The visibility of these structures is thought to be due to an excess of density relative to the ambient plasma, highlighting collections of magnetic field lines. The density enhancements are possibly due to local heating events associated with the magnetic fields \citep[e.g.,][]{CAR1994,Yokoyama_2001,Gudiksen_2005,Klimchuk_2008}.

Past efforts have placed significant attention upon understanding the influence of an inhomogeneous plasma on Alfv\'enic wave propagation. However, this has, until recently, been largely split into distinct areas of focus. One of these areas was predominantly interested in modelling the impact of Alfv\'enic waves on the solar wind \citep[for reviews see, e.g.,][]{BRUCAR2005,Cranmer_2017}, and hence focused on the inhomogeneity along the magnetic field, though a number of studies attempt to include the influence of the lower solar atmosphere as well. However, others were particularly interested in the role of perpendicular structuring, driven primarily by interests in wave heating of coronal loops \citep[for reviews see, e.g.,][]{Arregui_2015, 2020ARA&A..58..441N}. The final camp focused on wave propagation through partially ionized plasmas, investigating the role of ion-neutral effects \citep[for a review, see, e.g.,][]{Ballester_2018}. The division into these areas is, in part,  due to the vast range of scales that would be required to incorporate all the vital physics from the photosphere to the corona (and beyond). However modern numerical resources are enabling these areas to draw closer together. {We should mention that a number of works that have sought to bring together inhomogeneity parallel and perpendicular to the magnetic field. The studies where largely focused on emulating and predicting properties of strongly damped resonant coronal loop oscillations \citep[starting with, e.g.,][]{ANDetal2005b,ANDetal2005a,VERTH2007,VERTHetal2007}, and the opportunities for magneto-seismology \citep[see, e.g., ][for related reviews]{ANDetal2009,Nakariakov_2021}. }

\medskip

In the following we aim to provide an overview on the role inhomogeneities play in Alfv\'enic wave propagation. The phenomenon we discuss are likely to be present across the solar atmosphere, although the specifics (e.g., associated time-scales, magnitudes) of the wave generation, propagation and dissipation will depend upon the local plasma conditions. However, to bring together the details, it is necessary to outline a stage upon which wave propagation and evolution takes place. Here we will focus on wave propagation in the quiescent Sun and coronal holes, and largely avoid active regions. This choice is mainly driven by the fact that the magnetic structure and plasma conditions in the quiescent Sun and coronal holes are similar, at least in the lower solar atmosphere. Further, active regions are relatively infrequent features on the Sun when compared to the quiet Sun. Hence, the description of wave propagation in the quiet Sun provides a more general picture. This is not to say that the details of Alfv\'enic waves in active regions are unimportant or uninteresting, far from it\footnote{There are also plenty of reviews that focus on waves in active regions \citep[see, e.g.,][for a recent review]{2020ARA&A..58..441N}.}. We also note that in some of the given references, the analytic and numerical results are tailored to plasma conditions thought to be representative of coronal loops in active regions. Hence, it should be kept in mind that the numerical values given in those papers with respect to the phenomenon under study will need to modified for the quiet Sun.

\medskip

\subsection{Basic structure of the solar atmosphere}

\begin{figure*}[!t]
\centering
\includegraphics[scale=0.5]{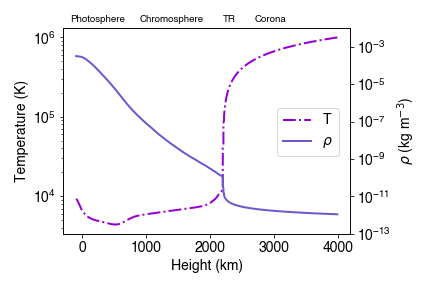}
\includegraphics[scale=0.5]{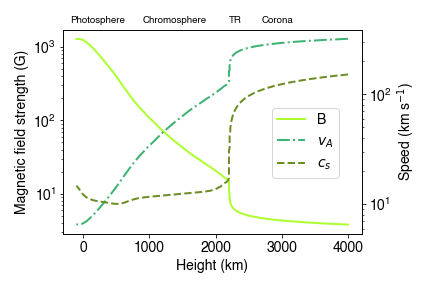}
\caption{A descriptive model of the solar atmosphere. The left panel shows temperature and density profiles. The lower solar atmosphere up to $\sim2000$~km is based on the FAL quiet Sun model (see text) and an additional coronal extension. The right hand panel displays an empirical magnetic field model for a magnetic bright point with a photospheric field strength of 1300~G. The corresponding Alfv\'en and sound speeds for the model atmosphere are also shown.}\label{fig:T_rho_atmosphere}
\end{figure*}

To highlight some of the key inhomogeneities that affect the Alfv\'enic waves, it is worth providing a description of the basic structure of the Sun's atmosphere. The left panel of Figure~\ref{fig:T_rho_atmosphere} shows the temperature and mass density profiles as a function of height in the lower solar atmosphere, from photosphere to the coronal base. The atmospheric profile shown is from a series of models by \citet{Fontenla_1993}, of which we select the quiet Sun profile (referred to as the FAL-C model). The FAL model is extended into the corona following \citet{2017ApJ...840...20S} with an assumption of a fully ionised corona\footnote{The data and source code for generating the atmospheric profiles shown throughout this review are available at: \url{https://github.com/Richardjmorton/FAL_models}.}.

As noted by \citet{Rutten_2021}, it is worth recognising that the FAL models really only represent the atmosphere of a solar-analog. 
The atmosphere in these models is 1D plane parallel with no magnetic field and lacking any of the dynamics that are observed (e.g. waves, shocks, granulation, spicules). However, the FAL models \cite[and the earlier VAL versions][]{Vernazza_et_al_1981} are a vital starting point for many investigations into wave propagation through the atmosphere of a Sun-like star. With this in mind, it can be seen that the 
temperature and density are expected to continuously vary with height. To complement the temperature and density, Figure~\ref{fig:T_rho_atmosphere} also shows an 
empirical model for the variation in vertical magnetic field strength \citep{Leake_2005}. The shown magnetic field profile should represent the expanding magnetic field emanating from intense magnetic flux tubes embedded in the photosphere \citep[known as magnetic bright points due to their appearance in certain photospheric diagnostics, e.g.,][]{BERTIT1996,Tsuneta_2008, DEWetal2009,Ito_2010}. In combination with the mass density, one can 
estimate the Alfv\'en speed ($v_A=B/\sqrt{\mu_0\rho}$) profile through the atmosphere (and using the prescribed temperature profile we 
also provide the sound speed, $c_s$, profile).

For the Alfv\'enic wave propagation, the temperature structure is usually not a major consideration due to the highly incompressible nature of the waves (although will play a role in the story, e.g., through its influence on the ionisation state of the plasma - Section~\ref{sec:sub_PI}). Of particular importance is the mass density, and to a degree the magnetic field, which shapes the details of the waves' journey through different layers of the stratified solar atmosphere. It can be seen in Figure~\ref{fig:T_rho_atmosphere} that the mass density has a scale height of 100-200~km in the lower solar atmosphere, which leads to a rapid increase in Alfv\'en speed. To conserve wave energy flux ($F=\frac{1}{2}\rho v^2v_A$), this leads to wave amplification, which can then induce non-linear mode coupling and shocks 
(discussed in Section~\ref{sec:chrom}). Upwardly propagating waves then encounter the sharp change in density that occurs at the transition region, which acts as a frequency filter and is a decisive factor in determining the Alfv\'enic wave energy that is able to enter the corona. The change in Alfv\'en speed in the corona and above is much more gradual, although it no less interesting; it is a key factor in the development of turbulence (Section~\ref{sec:sub_turb}).

\medskip

\begin{figure}[!t]
\centering
\includegraphics[scale=0.18]{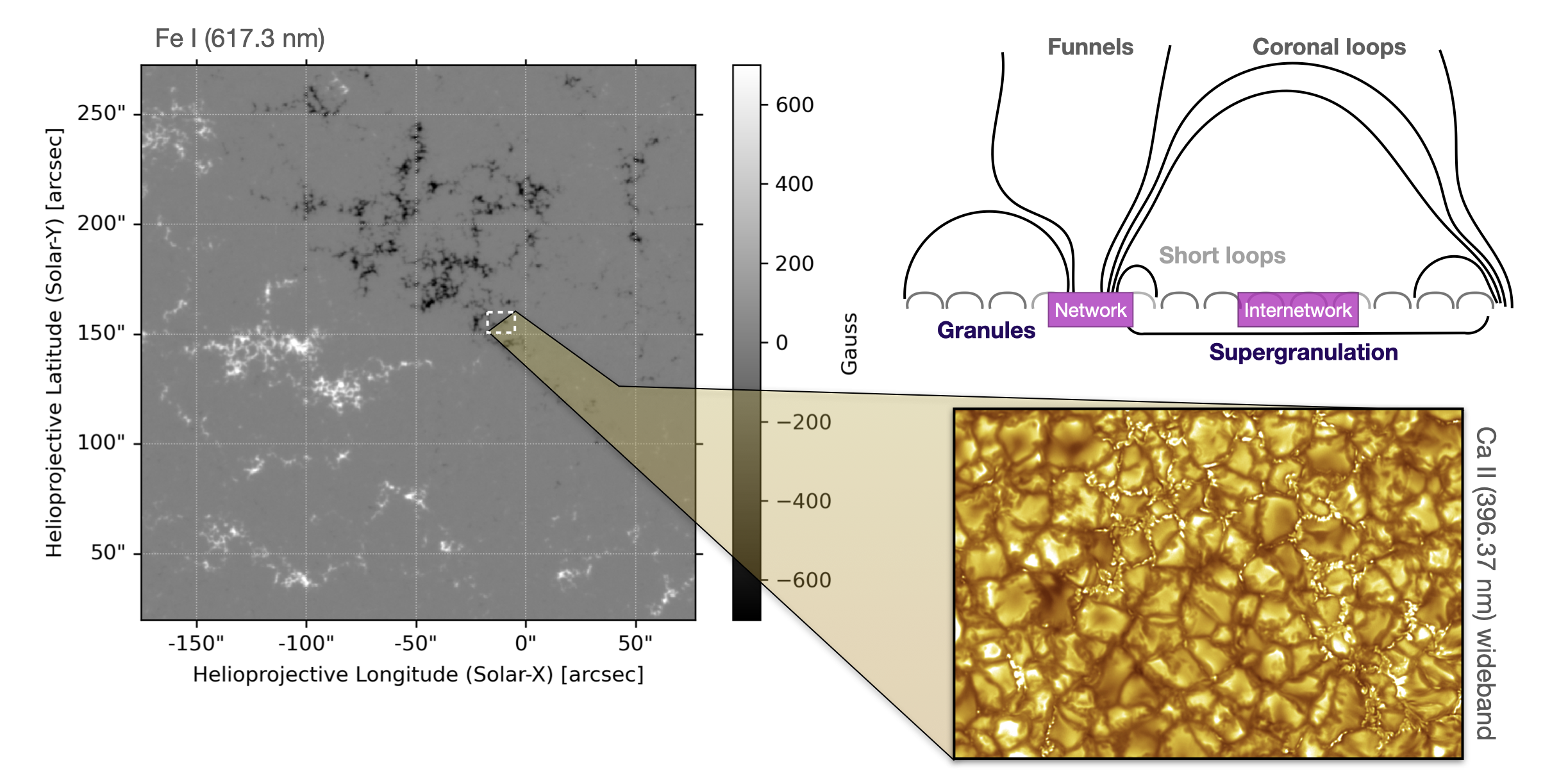}
\caption{A sketch of the organisation of the magnetic field in the quiet Sun. The left hand image is the photospheric magnetic field 
measured from the Zeeman splitting on an Fe I line (a magnetogram) by the Heliosesmic \& Magnetic Imager onboard SDO. The central feature of the image is relatively large clusters of magnetic field that highlight the network. The bottom right panel shows a small portion of the network, revealing the small-scale magnetic flux features as bright points (image courtesy of V. Henriques, taken with the Swedish Solar Telescope operated by Institute for Solar Physics, Stockholm, Sweden).   }\label{fig:mag_field}
\end{figure}

\subsection{The large-scale organisation of the magnetic field}

The large-scale organisation of the magnetic field plays an important role in characterising the spatial and temporal evolution of waves in the solar atmosphere. Hence, we believe it is important to overview critical aspects of this element here. A complete assessment, however, will require many more pages though we direct an avid reader to many other excellent texts discuss the structure, dynamics and evolution, \citep[e.g.,][]{schrijver_zwaan_2000,DEWetal2009,2019LRSP...16....1B}.

The organisation of the magnetic field depends upon height in the solar atmosphere, but is ultimately engendered by its distribution in the photosphere. The gas density and temperature vary by orders of magnitude between the photosphere and the corona, with the dynamics of the plasma controlling the evolution of the field in the photosphere (ratio of plasma pressure to magnetic pressure, or plasma beta, being greater than 1). Eventually, the rapid decrease in plasma pressure means that the magnetic field begins to dominate both the appearance and dynamics in the upper chromosphere and beyond (where plasma beta is less than 1). 

At the photospheric level, the magnetic field is found to emerge as (or coalesce into) small, intense kilogauss elements concentrated in the lanes between the granules (these magnetic bright points are seen in the Ca II image in Figure~\ref{fig:mag_field}). Over time, many of these small elements are advected to the edges of the so-called network \citep{Zirin_1985, Orozco_Su_rez_2012, GOSetal2014}, which is thought to highlight the lanes of a super-granular flow pattern deep in the interior (Figure~\ref{fig:mag_field}). This produces a large-scale ($\sim15-30$~Mm) patchwork of magnetic field \citep{NOVSIM1988, 2004ApJ...616.1242D}, although is still made up from clusters of small-scale magnetic elements (see the magnetogram in Figure~\ref{fig:mag_field}). The small-scale elements within the interior of the network (or internetwork) are generally thought to be the foot-points of low-lying, closed magnetic fields, extending only as high as the chromosphere and transition region \citep{Mart_nez_Gonz_lez_2009,WIEetal2010}. In contrast, those magnetic fields at the network boundaries extend into the upper atmosphere, spreading out as the gas pressure falls off and eventually filling the volume of the corona. In the upper right panel of Figure~\ref{fig:mag_field} we sketch this scenario. These magnetic fields are then either closed, connecting with opposite polarity flux and spanning the internetwork region; or they are open funnels, connected to the solar wind \citep[more detailed discussions can be found in, e.g.,][]{PET2001, WEDetal2009}. The separation of network and internetwork fields is likely not so neat, with the potential for short loops to exist with a foot-point in both \citep{WIEetal2010} and internetwork fields also contributing to the funnels.

\subsection{The fine-scale structure}\label{sec:fs_struc}

Near the top of the chromosphere, the magnetic field begins to dominate both the appearance and dynamics of the solar atmosphere. As mentioned, the plasma beta is much less than one in these regions. This is complemented by high magnetic Reynold's number ($\gg1$), meaning that the magnetic field obeys the so called `\textit{frozen-in}' condition which prohibits the cross-field diffusion of the plasma. Heating events can lead to evaporation or injection of plasma into the upper parts of the atmosphere and the consequence of the `\textit{frozen-in}' condition is that the plasma is confined to move along the magnetic field lines. This leads to enhancements of density along bundles of magnetic field lines, and provides the fine-scale structure seen in many images of the chromosphere\footnote{It is suggested by \citet{RUTVAN2017} that some of the chromospheric fine structure observed in Hydrogen $\alpha$ images is actually a delayed signature of heating event that took place earlier. Hence, the observed features may not represent the current topography of the magnetic field. } and corona (e.g., Figure~\ref{fig:corona}). The fine-structure has many forms, from elongated chromospheric fibrils and coronal loops, to short and highly dynamic spicules that originate in the chromosphere. The fibrils, spicules \citep[e.g.,][]{RUT2007,TSIetal2012,JESetal2015,2019ARA&A..57..189C} and active region coronal loops \citep[e.g.,][]{Reale_2014} have been well-studied . However, many of the other density striations seen in Figure~\ref{fig:corona}, e.g., quiescent coronal loops and the near radial features in polar coronal holes, are presently little-studied.

In both the chromosphere and corona, the well-studied structures have widths down to the resolution of current instrumentation; with typical widths around $\sim200-300$~km \citep[e.g.][]{Antolin_2012,PERetal2012,BROetal2013,Morton_2014}. Estimates for the over-density of some of these features, namely coronal loops in active regions, has been possible. For example, \citet{ASCetal2000b} suggests the density of warm coronal loops is an order of magnitude more than that expected from hydrostatic equilibrium, which the ambient plasma is expected to be in. More recently, \citet{Pascoe_2017} use seismology and estimate over-density values in the range 1.5-5.

\begin{figure}
\includegraphics[trim=0 0 0px 0px, clip, scale=0.12]{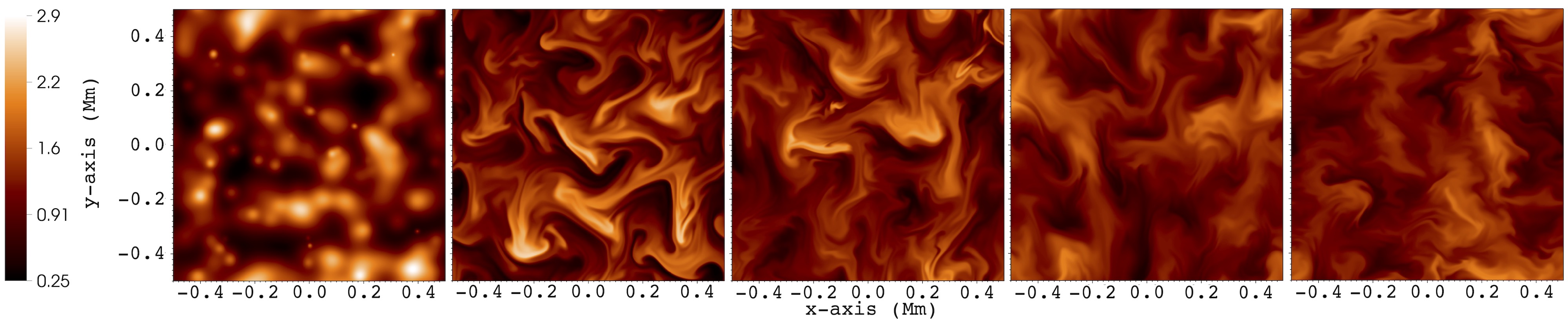}
\caption{The destruction of simplified loop geometries. The figure shows the cross-sections of coronal loops, displaying the density. The magnetic field is orientated perpendicular to the 2D slice shown. The loops are initially over-dense with a Gaussian density profile in the cross-section (left hand panel), which gradually becomes distorted over time. The time step between panels is 250~s and the density colour-bar is shown in units of $10^{-12}$~kg m$^{-3}$. Figure reproduced from \citet{Magyar_2017}. }\label{fig:Magyar_2017}
\end{figure}

Apart from the density and cross-sectional width of fine-scale structures, another noteworthy aspect is the cross-sectional geometry of the flux tubes in the solar atmosphere.
\citet{Parker_1974,Parker_1974b} suggested cylindrical geometry of these features using a simplistic hydrodynamic mechanism involving turbulent pumping to coalesce magnetic fields into the slender, tube-like structures. This simplified assumption remained a cornerstone to the wave studies in localised chromospheric and coronal structures \citep[e.g.,][]{Goossens_2011}. However, there has been some work that has ventured from this assumption, \citep[e.g.,][]{MORERD2009,PASetal2011,MORRUD2011,Guo_2020}. The observational evidence for the actual cross-sectional geometry of the chromospheric and coronal structures is scant. Theoretical work suggests that the shape of the flux tube cross section can vary along the structure, and is likely oblate \citep{RUD2009, Malanushenko_2013}. Observational studies are particularly complex, and appear to show mixed results \citep{McCarthy_2021}. Moreover, recent simulations have demonstrated that, even when starting with circular cross-sections, the presence of transverse waves lead to a distortion of the flux tubes cross-section through instabilities \citep[e.g.,][and discussed further in Section~\ref{sec:sub_insta}]{antolin2014,magyar2016,Magyar_2017,antolin2019}. Figure~\ref{fig:Magyar_2017} shows simulation results from \citet{Magyar_2017} that demonstrate this process in action. 

Furthermore, it has been speculated that the appearance of a near-circular tube-like geometry of the fine-scale structures could be due to observational artefacts. \citet{Judge_2011} suggested that some observed chromospheric spicular structures are wrapped current sheets that appear as thin, elongated features. More recently, \citet{2022ApJ...927....1M} put forward the idea that some bundles of coronal loops are optical illusions in EUV passbands. The authors suggest that these features are actually complex and diffuse objects (`wrinkles in the coronal veil') that appear as compact, slender structures due to artifacts that arise from the inherent integration along the line of sight over extended regions in optically thin plasma. The authors examined the coronal structures using highly realistic numerical simulations, and found that the cross-sections of many coronal structures appear to show a highly turbulent appearance (similar to the cross-sections from \citealp{Magyar_2017} shown in Figure~\ref{fig:Magyar_2017}). However, it is not discussed how or why the corona is structured in such away. 

\medskip
Importantly, while the actual cross-sectional shape of flux tubes is unknown, both observations and simulations suggest that there are plasma inhomogeneities in the direction perpendicular to the magnetic field. As we shall now discuss, this enables a wealth of interesting wave physics to take place and is likely a key factor in enabling the dissipation of wave energy in the chromosphere and corona.

\section{Linear Alfv\'en Wave Theory}\label{sec:wave_theory}
To appreciate the impact of an inhomogeneous plasma on wave propagation, we briefly discuss some of the fundamental theoretical aspects of linear Alfv\'enic waves in a structured plasma. It turns out that perpendicular inhomogeneity in the plasma modifies the classical picture of wave behaviour presented in the textbook examples of homogeneous plasma, and removes the clear division between Alfv\'en and magneto-sonic modes. {It was noted as early as 1982 by \cite{1982alwa.book.....H} that the presence of inhomogeneity leads to coupling of the total pressure to the dynamics for Alfv\'en modes.} Inhomogeneity also enables the MHD waves to have a different appearance in different parts of the plasma, readily modifying their dominant characteristics. The following aims to be a discussion of the salient points, and we recommend readers refer to the original texts for deeper discussions \citep[e.g.,][]{1982alwa.book.....H,2003ASSL..294.....G, GOOetal2012,2019FrASS...6...20G}. We first discuss the case of a homogeneous plasma before moving onto the more interesting physics that occurs when this restriction is dropped.

\subsection{The homogeneous case}
To begin, let us assume that we have a plasma with magnetic field that is oriented in the $z$ direction and is 
homogeneous in all plasma quantities along the field. We make no assumptions about the plasma perpendicular to the 
field at this moment. The linearised ideal MHD equations can be reduced to \cite[e.g.,][]{GOOetal2012},
\begin{eqnarray}
    \omega^2\xi_z-k_zc_s^2Y &=&0 , \label{eq:mag_son1}\\
    k^2v_A^2k_z\xi_z+(\omega^2-k^2(c_s^2+v_A^2))Y & = & 0\label{eq:mag_son2}\\
    (\omega^2-k_z^2v_A^2)Z &=& 0, \label{eq:alf}
\end{eqnarray}
for planar, harmonic waves, i.e., wave variables are proportional to $\exp(ik_zz-i\omega t)$, where $\omega$ is the 
frequency and $k_z$ is the longitudinal wave number. Here the equations have been expressed in terms of variables that 
expose the underlying physics. The variables are the magnitude of the wavevector $k=\sqrt{k_x^2+k_y^2+k_z^2}$, the 
sound speed $c_s$, the Alfv\'en speed $v_A$, and the Lagrangian displacement along the magnetic field, $\xi_z$. 
Additionally, we have also used the plasma compression,
\begin{equation}\label{eq:comp}
   Y=-i\nabla\cdot\vec{\xi}, 
\end{equation}
and the vorticity parallel to the magnetic field,
\begin{equation}\label{eq:par_vort}
    Z = -i(\nabla\times\vec{\xi})_z,
\end{equation}
where $\vec{\xi}$ is the Lagrangian displacement vector. It is also useful to define the ratio of the horizontal components of the total pressure force and the magnetic tension force \citep{GOOetal2012}, which is given by,
\begin{equation}\label{eq:force_ratio}
    \Lambda(\omega^2)=\frac{\omega^2}{(k_zv_A)^2}-1.
\end{equation}

\medskip
In an homogeneous media of infinite extent, the system of equations (\ref{eq:mag_son1})-(\ref{eq:alf}) are decoupled into two systems. One of these describes Alfv\'en waves and contains only the vorticity along the field, i.e., Eq.~\ref{eq:alf}. There is no compression or displacement along the magnetic field ($Y=0$, $\xi_z=0$). The solution to this set of equations results in the well known dispersion relation,
$$
\omega = k_zv_A.
$$
Hence, from Eq.~\ref{eq:force_ratio} it can be found that $\Lambda(\omega^2)=0$, and magnetic tension is the only restoring force. The other system involves the variables $\xi_z$ and $Y$, and defines the magneto-sonic modes, i.e., Eqs.~\ref{eq:mag_son1} and \ref{eq:mag_son2}. Unlike the Alfv\'en waves, these magneto-sonic modes are compressible and have a longitudinal displacement but do not propagate parallel vorticity.

\subsection{Perpendicular inhomogeneities}\label{sec:wt_in_perp}
The addition of an inhomogeneity in the direction perpendicular to the magnetic field leads to a coupling of the wave variables, which, in the case of a straight magnetic field, is mediated by Eulerian perturbation of the total pressure \cite[for a twisted field there is a stronger coupling arising from the so-called coupling functions, e.g., ][]{1991SoPh..133..227S,2019FrASS...6...20G}. {This coupling leads to MHD waves having mixed properties and is present even in the linear MHD system \citep[e.g., ][]{Goossens_2002}. } There are two scenarios that are worth discussing as they highlight key aspects of increasing the geometric complexity.

The first case of interest is two uniform plasma separated by a discontinuity in the Alfv\'en velocity, which we can assume arises due to a discontinuity in the density, i.e.,
$$
\rho(x) = \begin{cases}
\rho_i & \hspace{0.5cm} x\le x_0,\\
\rho_e & \hspace{0.5cm} x> x_0.
\end{cases}
$$
At such an interface, Alfv\'en surface modes are able to exist \citep{1979A&A....76...20W,1979ApJ...227..319W,1981SoPh...69...27R}. In the limit of `nearly perpendicular propagation' ($k_y^2>>k^2_z$), the dispersion relation for the surface mode is given by:
\begin{equation}\label{eq:kink}
\omega^2 = k_z^2\frac{\rho_iv^2_{Ai}+\rho_ev^2_{Ae}}{\rho_i+\rho_e}\equiv \omega_k^2,
\end{equation}
which lies between the Alfv\'en frequencies for both plasmas. Hence, $\Lambda$ (given in Eq.~\ref{eq:force_ratio}) is small on either side of the discontinuity and magnetic tension dominates the restoring forces. The surface mode is also largely insensitive to the value of the sound speed. Furthermore, it has the property that parallel vorticity is zero everywhere except at the discontinuity, contrary to the classical Alfv\'en wave in uniform plasma \citep{GOOetal2012}. For such a mode, an appropriate term is Alfv\'enic\footnote{{The term Alfv\'enic was already used by \cite{ION1978} to describe the surface Alfv\'en wave.}}, which highlights such modes are different from the classical Aflv\'en wave, and has mixed properties.

\medskip

The second case is that of a 1D magnetic cylinder, which we describe in cylindrical coordinates, ($r$, $\theta$, $z$). The magnetic field is still oriented in the $z$ direction.
Since the background plasma depends only on perpendicular inhomogeneities, which are in the radial ($r$) direction, then the perturbations, independent of azimuth ($\theta$) and $z$, are proportional to
$$
\exp(i(m\theta+k_zz-\omega t)),
$$
where $m$ is the azimuthal wavenumber. In this case, Eqs.~\ref{eq:comp} and \ref{eq:par_vort} become
$$
Y = \frac{1}{r}\frac{\partial (r\xi_r)}{\partial r}+im\xi_\theta+ik_z\xi_z, \hspace{1cm} Z=\frac{1}{r}\frac{\partial (r\xi_\theta)}{\partial r}-im\frac{\xi_r}{r}.
$$
For the case of $m=0$, it can be seen from these expressions for $Y$ and $Z$ that Eqs.~\ref{eq:mag_son1}-\ref{eq:alf} are decoupled. The solutions to the two sets of equations are axisymmetric modes, namely the torsional Alfv\'en mode and the sausage magneto-sonic mode. The Alfv\'en mode depends only upon $\xi_\theta$, as each magnetic surface oscillates at its own frequency. As noted by \citet{2019FrASS...6...20G}, this is the only pure Alfv\'en mode in a non-uniform 1D cylindrical plasma. {The nature of the coupling is further elucidated by the coupling function derived by \cite{1991SoPh..133..227S} for a straight magnetic field, namely,
\begin{equation}
    C_A=\frac{m}{r}B_zP',
\end{equation}
where $P'$ is the total pressure perturbation. It can be seen that a total pressure perturbation is also required to be non-zero in order for coupling, as well as $m\neq0$.}

Under the simplifying assumption of a magnetic flux tube with a piece-wise constant density \cite[e.g.,][]{SPR1982,EDWROB1983}, solutions exist for any $m$ that can be identified with Alfv\'en waves (i.e., solutions with $\nabla\cdot\vec{v}=0$). More generally, for any geometry, a pure Alfv\'en mode can exist if there is an ignorable transverse direction.
However, in the non-uniform 1D cylinder, modes with $m\neq0$ have mixed properties and all wave variables are coupled. {All eigenmodes of the system now have all perturbed quantities that are non-zero \citep{Goossens_2002,2019FrASS...6...20G}.} Here, the modes cannot be separated by the distinctions found in the homogeneous case. Also of interest, it turns out all wave modes in inhomogeneous plasma are efficient generators of vorticity \citep{2019FrASS...6...20G}.

\medskip

Before moving on, it is worth mentioning the non-axisymmetric kink mode (when $m=1$) in detail. This particular wave mode has been the subject of great interest over the last two decades. Transverse
oscillations of coronal loops, first noticed in observations back in the 1990's, were interpreted as 
the kink mode \citep{ASCetal1999,NAKetal1999}. Then, after observations of transverse waves in the 
chromosphere \citep{DEPetal2007,OKAetal2007} and corona \citep{TOMetal2007} in 2007, the kink mode was subject to an 
intense discussion about the wave properties and nomenclature. In one of the seminal papers on MHD waves in 
flux tubes, \citet{EDWROB1983} used the prefix `fast' to describe the mode. As mentioned, the solar atmosphere is highly inhomogeneous and MHD waves in this environment portray mixed 
properties and it is misleading for the modes to be classified using the labels derived from the homogeneous 
case \citep{2019FrASS...6...20G}. 

The characteristics of the kink mode were examined and discussed in detail by \citet{GOOetal2009}, where 
they demonstrated that the kink mode is in fact very robust and cares little for the plasma environment when $k_zR<1$ (here, $R$ is the flux tube's radius). Moreover, the kink mode was shown to be highly 
incompressible, with negligible pressure perturbations. The dispersion relation for the kink mode in a cylinder with a piece-wise constant density is given by Eq.~\ref{eq:kink}, while the ratio given by Eq.~\ref{eq:force_ratio} is
$$
\Lambda_i(\omega)=-\Lambda_e(\omega)=\frac{\rho_i-\rho_e}{\rho_i+\rho_e},
$$
in the interior and exterior of the cylinder. Hence, for any value of density contrast, this ratio is always less than one and magnetic tension is the dominant restoring force. This led Goossens at al. to suggest the descriptor for these waves should be Alfv\'enic, alluding to the dominant properties being close to classical Alfv\'en wave. In a follow up work, \citet{GOOetal2012} go onto to demonstrate that the fundamental radial modes for any non-axisymmetric wave mode ($m>0$) can be described as a surface Alfv\'en wave. In this case, the waves display the same properties as a surface Alfv\'en wave at a true discontinuity\footnote{Other radial overtones have properties that \citet{GOOetal2012} call fast-like.}. However, if the discontinuity is replaced by a region of non-uniform, continuously varying density, then the vorticity is present throughout this region and the modes are Alfv\'enic global/quasi modes\citep{Goossens_2002}. The presence of vorticity and the continuously varying density profile leads to a number of interesting effects that impact wave propagation, damping, and dissipation; which are discussed in section~\ref{sec:diss}.

\section{Wave excitation}\label{sec:wave_exci}

Recent theoretical models have shown that the inhomogeneous nature of the Sun's atmosphere could also play a key role in exciting Alfv\'enic waves. We first overview the existing paradigm surrounding the excitation of Alfv\'enic waves and then discuss mode conversion from inhomogeneities.

\subsection{Convection}

The typical picture of Alfv\'enic wave generation begins at the photosphere and was suggested many decades ago \citep[e.g.,][]{OST1961}. The turbulent patterns of granulation are the most common feature of the photospheric topology 
(Figure~\ref{fig:mag_field}), that manifest from the convective motions deep within the solar interior. The average lifetime of granular cells is 5-10 minutes and have horizontal 
scales of about 1~Mm \citep[e.g.,][]{1997ApJ...475..328S}. As mentioned in Section~\ref{sec1}, the boundaries of these granular cells appear as dark lanes where strong plasma downflows are observed. These intergranular lanes form favourable sites for harbouring intense concentrations of magnetic flux known as magnetic bright points (MBPs). The MBPs are seen to be passively advected with the large-scale super-granule flow but are also subject to buffeting from the granulation \citep{BERTIT1996,BERetal1998,VANBALLetal1998,NISetal2003,CHITetal2012}. The associated horizontal motions are turbulent and rapid, with an average root-mean-squared 
(RMS) velocity of $\sim1-1.5$~km/s and correlation times of 20-30~s \citep{CHITetal2012}, while the line-of-sight (LOS) Doppler velocity magnitudes are of the order $\sim$1.2~km/s \citep{Nordlund_2009,Oba_2017}. 

These motions correspond to the both horizontal and vertical shaking of the magnetic bright points and 
can induce various types of MHD wave-like fluctuations at the foot-points of thin magnetic flux tube
structures \citep{1981A&A....98..155S,1998ApJ...495..468S}. While vertical motions can drive sausage or 
longitudinal modes \citep{Defouw_1976,ROBWEB1978}, the horizontal motions can induce transverse waves, 
namely the kink mode, if using a flux tube description for the bright points \citep{1981A&A....98..155S,CHOetal1993,2002A&A...386..606M}. \citet{CRAVAN2005} suggested such kink
modes would convert to `classical' Alfv\'en waves in the low chromosphere when the individual magnetic 
elements expand and merge
with one another. This implies that the upper atmosphere is homogeneous perpendicular to the magnetic field (as is the case of the Cranmer \& van Ballegooijen model). However, as discussed in the introduction, the chromosphere and corona apparently possess a significant amount of inhomogeneity. Hence, the wave energy should propagate to upper layers as Alfv\'enic modes instead - although we are unaware of any
model/theory that addresses the details of how this may happen.  

Closely related to the horizontal motions are photospheric vortices, which are convective down-drafts in the intergranular lanes. 
They were initially detected through the observed motions of MBPs \citep{BONetal2008} and have been found to be ubiquitous across the photosphere \citep[e.g.,][]{WEDROU2009,MORetal2013,2018ApJ...869..169G,2019NatCo..10.3504L} and in numerical simulations \citep[e.g.,][]{SHEetal2011,MOLetal2011}. 
Such vortices appear to be capable of exciting rotational Alfv\'enic modes in numerical simulations \citep{2021A&A...649A.121B,Breu_2022},
and have also been suggested to enable distortions of flux elements shape and intermixing of field lines, leading to MHD wave excitation \citep{VANBALLetal2011}.

\subsection{Mode conversion}
\begin{figure}[!t]
    \centering
    \includegraphics[trim=0 0 70px 0px, clip, scale=3]{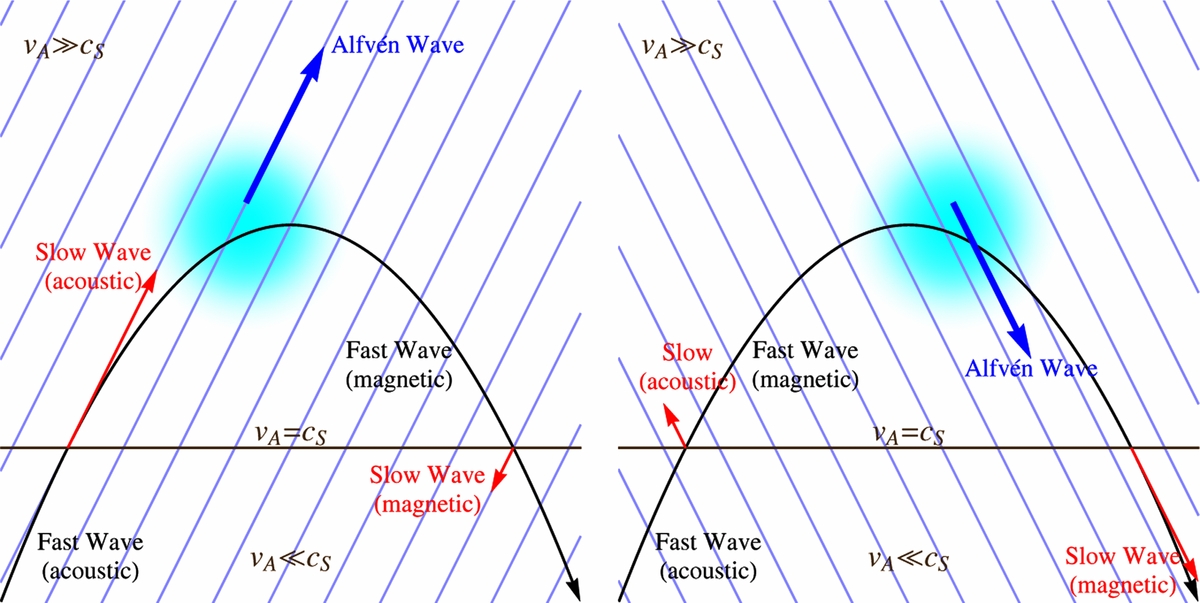}
    \caption{A sketch of the double mode conversion in the lower solar atmosphere that ends in the production of Alfv\'enic waves. The details are discussed in the text. The blue diagonal lines represent the magnetic field. Reproduced from \citet{KHOCAL2012} permission of the AAS.}
    \label{fig:koh_cal}
\end{figure}
More recently, there has been interest in alternative pathways for convective energy to reach the corona as Alfv\'enic waves. In part this interest has been driven by some fundamental theoretical results \citep{CALGOO2008}, but also by the apparent high reflectivity of the transition region to photospheric driven Alfv\'enic waves \citep[suggested to be $\sim95$~\% reflection rate by][]{CRAVAN2005}. The essence of this pathway is shown in Figure~\ref{fig:koh_cal}. 

It is well established that the Sun's interior (and the interior of many other stars with convective envelopes) posses a wide range of acoustic oscillations, known as \textit{p}(pressure)-modes. The \textit{p}-modes are largely confined to the solar interior but can leak out into the atmosphere. The acoustic modes can be absorbed by the magnetic fields of sunspots \citep{Bogdan_1993} and convert to magneto-acoustic modes \citep[e.g.,][]{Spruit_1992,Cally_1997}. This conversion occurs in the region of the equipartition layer (where $c_s=v_A$) and requires the waves to have frequency greater than the acoustic cut-off frequency. Here, field inclination also proves an important ingredient in enabling \textit{p}-mode absorption \citep{Cally_2000,Crouch_2003}. The inclination modifies the acoustic cut-off frequency \citep{BELLER1977} which enables lower frequency acoustic modes to propagate into the atmosphere (termed the `ramp' effect by Cally) and reach the equipartition layer. The variation of acoustic cut-off frequency with magnetic field inclination has been observed in sunspots numerous times \citep[e.g.,][]{McIntosh_2006,LOHetal2016,MORetal2021}. 

The absorption of \textit{p}-modes has also been demonstrated for slender magnetic flux tubes \citep{Bogdan_1996}, with similar mode transmission/conversion processes expected to take place \citep{BOGetal2003}. Observations by \citet{JEFetal2006} revealed the presence of low-frequency magneto-acoustic modes at supergranule boundaries indicates the ramp effect is also in action, with the authors terming the network magnetic fields as `magneto-acoustic portals'. 

\medskip
For those waves reaching the equipartition layer, the initial fast acoustic mode is converted to a slow mode; sometimes referred to as transmission as the energy of the wave maintains its basic character. This conversion is largely favoured in regions of vertical magnetic field, where the attack angle between the wave vector and magnetic field is small. However, when the attack angle is large, the \textit{p}-modes predominantly get converted to fast magneto-acoustic waves, which have a magnetic character. Due to the significant increase in the Alfv\'en speed in the upper chromosphere (Figure~\ref{fig:T_rho_atmosphere}), the fast wave is then refracted. Subsequently, around the height of refraction, another linear mode conversion takes places. This conversion is from fast to Alfv\'en, which is inherently 3 dimensional and requires the wavevector to be in a different plane to the magnetic field lines \citep{CALGOO2008,CALHAN2011,HANCAL2012,KHOCAL2012}. This pathway appears to enable a significant fraction of the \textit{p}-mode energy to reach the corona as Alfv\'enic waves \citep{KHOCAL2012}, with \citet{HANCAL2012} estimating that around 30~\% of the \textit{p}-mode flux gets carried through the transition region as Alfv\'enic waves. \citet{HANCAL2012} also provide a crude estimate for the energy flux in the 3-5~mHz band of 800~Wm$^{-2}$, which would mean that this process could provide a substantial contribution to meeting the energy requirements for maintaining a hot corona. 

While the just referenced works focus on plasma configurations with no perpendicular structuring, the influence of a inhomogeneous corona is investigated in \citet{CAL2017} and \citet{Khomenko_2019}. It is found that an incident fast wave on a collection of flux tubes largely converts to Alfv\'en waves. The incident wave is also subject to scattering in Fourier space, exciting the kink mode and fast waves in the corona.

At present, the observational evidence for the generation of coronal Alfv\'enic waves by mode conversion is still tentative. It has been found that the power spectra of coronal Doppler velocity fluctuations, interpreted as the kink mode, presents an excess of power at around $\sim3-4$~mHz \citep{TOMetal2007,MORetal2016,MORetal2019}. This excess power appears to be inconsistent with expectations from driving by horizontal granular motions, and the location in frequency space is coincident with the peak in \textit{p}-mode power. \citet{MORetal2019} demonstrated this enhanced power is present throughout the corona and also seemingly present through various stages of the Sun's magnetic activity cycle, implying the process is global and also somewhat insensitive to the global magnetic field structure. {Recent one dimensional simulations have attempted to incorporate the role of longitudinal waves in wave-driven solar wind models \citep{Shimizu_2022}. The results show an enhanced Alfv\'en wave flux in the corona leading to greater mass loss rates. While \citealp{Shimizu_2022} make a case for a linear mode conversion being responsible for the coronal Alfv\'enic waves in the model. However, it is not evident (to us at least) that the physical process in action in the model is a linear one, given the requirements for three dimensions expressed in previous works \citep[e.g., ][]{CALGOO2008}. Further, the result that the mode conversion rate is dependent upon amplitude \citep[Section 3.3 in ][]{Shimizu_2022} suggests that the process may be non-linear. Regardless, of the mechanism the work support a relationship between \textit{p}-modes and the observed coronal Alfv\'enic waves.}

\section{Wave damping}\label{sec:diss}

There are a number of mechanisms for energy exchange between the MHD waves, other wave modes, and the ambient medium. In a uniform plasma, the damping of Alfv\'en waves occurs on time scale $\tau_{d}={\lambda^{2}}/\eta$, where $\eta$ is the diffusivity and on length scale $L_{d}=v_{A} \tau_{d}$. However this dissipation is weak due to magnetic Reynolds number, $R_{m}$, being large, i.e., $R_{m} ={L_{d}}/{\lambda} \gg 1$. Theory shows that the presence of inhomogenity introduces additional mechanisms that enable the waves to damp more effectively. These damping/dissipation processes are strongly influenced by the inhomogeneity of the plasma and background magnetic field conditions, and act with varying levels of efficacy in transferring Alfv\'enic wave energy to the plasma (or other modes or scales). 
Perhaps unsurprisingly, it turns out that the interplay of multiple processes influences the efficacy of each mechanism in the solar atmosphere.

\subsection{Phase mixing and Resonant absorption}

The most well-known process for damping Alfv\'enic waves is probably phase mixing \citep{HEYPRI1983}, which requires a gradient in the Alfv\'en speed perpendicular to the direction of wave propagation. The Alfv\'en waves propagating along individual magnetic field lines (or magnetic surfaces) become out of phase as they propagate, generating small scales and thus increasing local gradients perpendicular to the field. The previously discussed equations for Alfv\'enic waves, i.e., Eqs.~\ref{eq:mag_son1}-\ref{eq:alf}, are based on the ideal MHD equations. \citet{HEYPRI1983} demonstrated that including viscosity ($\nu$) and magnetic diffusivity ($\eta$) leads to an equation for shear Alfv\'en waves (under some restrictions) of the form,
\begin{equation}
    \frac{\partial^2 v}{\partial t^2}=v_A^2(x) \frac{\partial^2 v}{\partial z^2}+(\nu+\eta) \frac{\partial^2 }{\partial x^2} \frac{\partial v}{\partial t},
\end{equation}
where $v$ is the velocity perturbation and the magnetic field is oriented in the $\hat{z}$ direction. It is clear from this equation that the increase in gradients due to phase mixing enhances the viscous and Ohmic dissipation of the wave energy. The length-scale over which phase mixing occurs is given by \citep[e.g.,][]{Mann_1995},
\begin{equation}
 L_{ph} = \frac{2\pi}{t|d\omega_A(x)/dx|},
 \label{eq:L_pm}   
\end{equation}
and can be seen to depend on the gradients of the variation in Alfv\'en speed. Here $\omega_A\approx v_Ak_\parallel$, with $k_\parallel$ the wavenumber parallel to the magnetic field.

\medskip

Closely related to phase mixing is resonant absorption \citep{ION1978}. If we assume that some region of plasma has a continuous variation in density or magnetic field, then the Alfv\'en speed is a continuous quantity and there exists a continuum of Alfv\'en modes with $\omega_A(x)$. If Alfv\'en waves are driven with a frequency, $\omega_D$, that lies within the continuum, then a resonance occurs at locations where $\omega_A(x)=\omega_D$. Hence, wave energy can be transferred from large-scale motions to localised, small-scale motions. This is also true for surface Alfv\'en waves, where the discontinuous region discussed in Section~\ref{sec:wt_in_perp} is replaced by a region with continuously varying Alfv\'en speed \citep{Lee_1986,HOLYAN1988}. As with phase mixing, the inclusion of viscous and Ohmic dissipation enables the wave energy at the resonance to be dissipated effectively. {Interestingly, as well as enabling the coupling of the wave variables (Sections~\ref{sec:wave_theory}, \ref{sec:wt_in_perp}), the total pressure plays an important role in resonant absorption. \cite{HOLYAN1988} noted that \textit{`[resonant] absorption can occur in any situation where the total pressure perturbations are imparted to field lines satisfying the Alfv\'en ... resonance conditions'.} Moreover, resonant absorption is able to modify the nature of classical Alfv\'en waves, introducing non-zero total pressure variations and compressibility \citep{Poedts_1989,Poedts_1990,Goossens_1992}. }  

Due to the appearance of localised flux tubes in
images of the corona (e.g., Figure~\ref{fig:corona}), many studies in the last two decades 
have focused on examining wave propagation in a cylindrical wave guide with a continuous density profile perpendicular to the magnetic field
\citep[see, e.g.,][for reviews]{Goossens_2005,Goossens_2011}. The region of continuous 
density has often been limited to a narrow annulus at the boundary of the wave guide\footnote{Mainly for the sake of mathematical simplicity.}, and referred to as the inhomogeneous layer. In the case of the kink mode, the kink frequency lies in the Alfv\'en continuum, i.e. $\omega_{Ai}<\omega_k<\omega_{Ae}$
and the global energy of the kink mode is transferred to other Alfv\'enic modes in the inhomogeneous layer \citep[rotational modes in the case of the cylinder, e.g.,][]{Poedts_1989,1991SoPh..133..227S,PASetal2012,GOOetal2013,Giagkiozis_2016}. Similar to phase mixing, the rate of resonant absorption is found to depend on the degree of the inhomogeneity. For propagating kink waves\footnote{A similar expression for damping rate is also obtained for standing modes \citet{RUDROB2002}.} it was shown by \citet{TERetal2010c} that, under the approximation of thin tube and thin boundary, the damping length of the kink mode due to resonant absorption is given by:
\begin{equation}
    L_D = \lambda R \frac{4}{\pi^2}\frac{\rho_i+\rho_e}{(\rho_i-\rho_e)^2}\left|\frac{d\rho}{dr}\right|_{r_A},
\end{equation}
where $R$ is the flux tube radius and the term $\left|{d\rho}/{dr}\right|_{r_A}$ is the gradient of density across the resonant layer. This expression clearly reveals that the density difference between the wave guide and the ambient plasma is inversely proportional to the damping length, which results in a higher rate of resonant absorption when the inhomogeneity contrast is larger. In terms of observed wave damping in the solar atmosphere, direct evidence for resonant absorption is ambiguous. However, the theory appears to give a good account of the observed damping of standing \citep[e.g.,][]{Aschwanden_2003,VERetal2013b} and propagating kinks modes \citep{VERTHetal2010,TIWARI_2019,tiwari_2021,Morton_2021}. 

\medskip 

\begin{figure}[!t]
    \centering
    \includegraphics[trim=0 2.7cm 0 0, clip, scale=2.5]{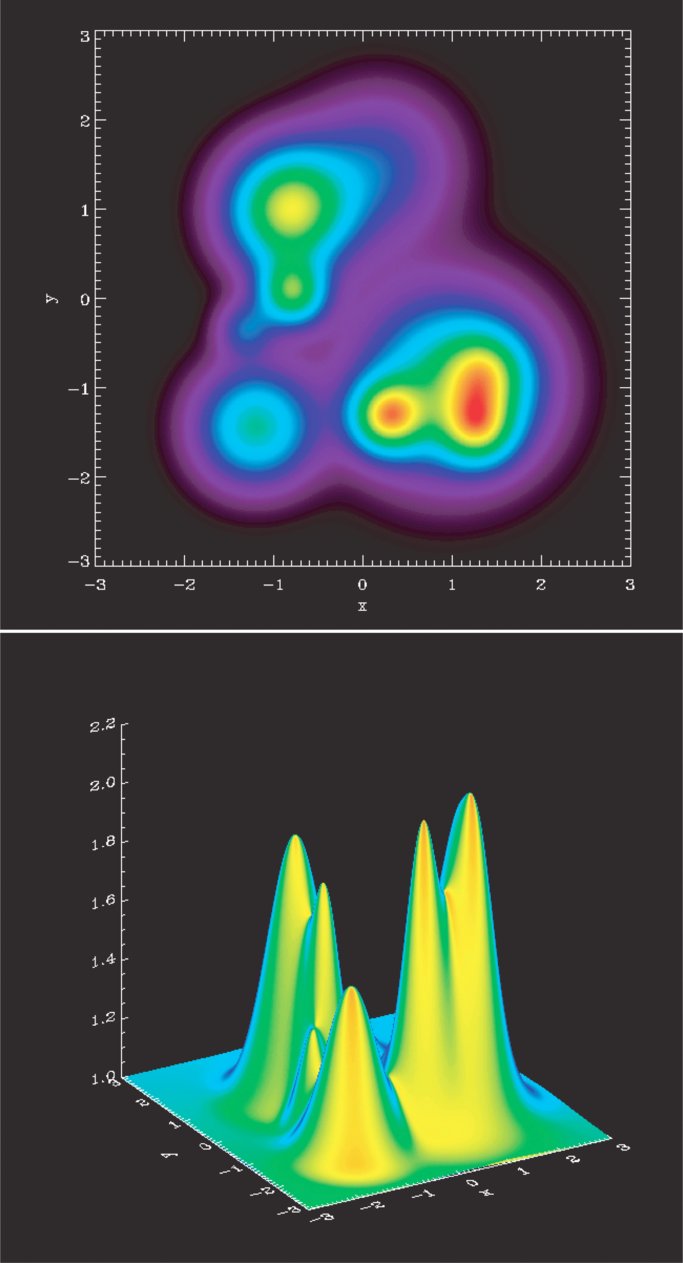}
     \includegraphics[trim=0 0 0 2.7cm, clip,scale=2.5]{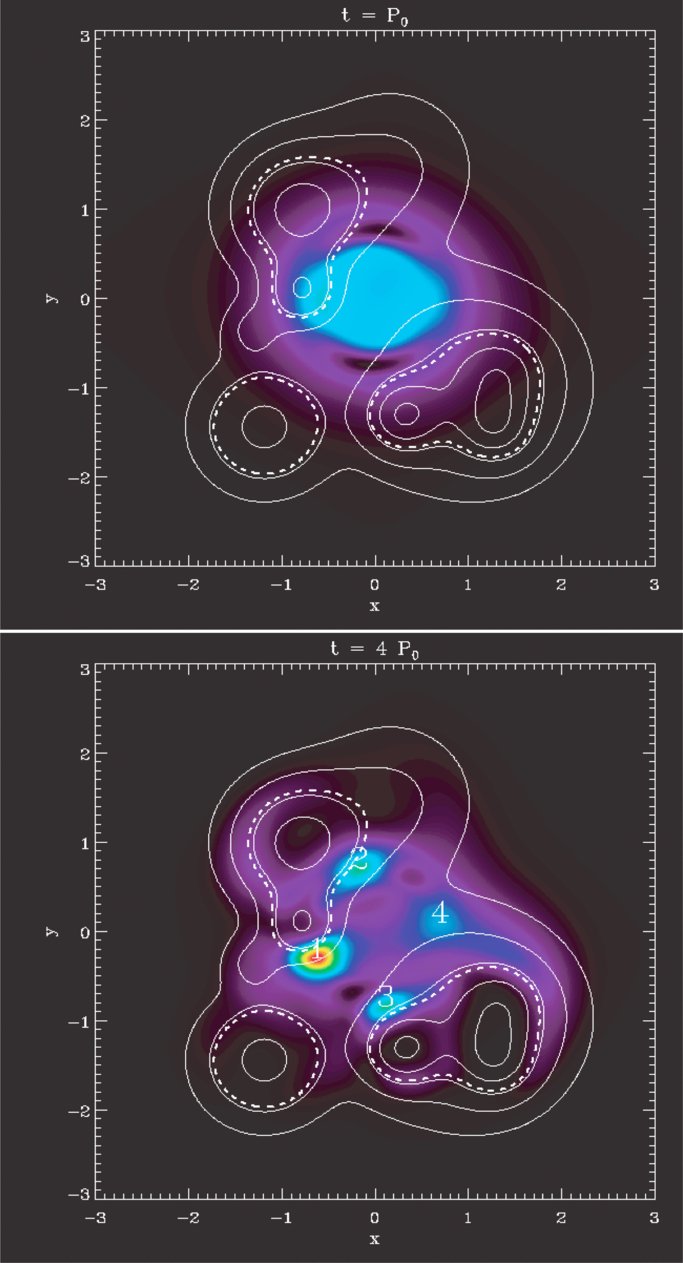}
    \caption{An example of resonant absorption in an inhomogenous corona. The left panel shows the contours of density in a cross-section perpendicular to the magnetic field. The largest density is a factor of 2 greater than the external plasma. The right panel shows the concentration of wave energy at regions where the local Alfv\'en speed equals the kink speed of the local density maxima. Figure adapted from \citet{PASetal2011} with permission of the AAS.}
    \label{fig:pascoe_2012}
\end{figure}

The presence of a continuous Alfv\'en speed profile showcases the 
first example of the joint influence of different wave damping mechanisms; in this case, resonant absorption and phase mixing. The resonant Alfv\'enic waves within the inhomogeneous layer (arising from the transfer of energy from the kink mode) can then propagate along magnetic surfaces at different Alfv\'en speeds, hence are subject to phase mixing \citep{Terradas_2008,SOLetal2015}. The expression for phase mixing lengths, Eq.~\ref{eq:L_pm}, has been shown to provide a good description of the development of small-scales in analytical and numerical models of kink mode damping \citep[e.g.,][]{PASetal2013,SOLetal2015}. This dual process has been demonstrated to be robust in multi-stranded structures \citep{Terradas_2008} and with arbitrary inhomogeneities \citep{PASetal2011}. Figure~\ref{fig:pascoe_2012} shows numerical results from \citet{PASetal2011}, where a  collection of over-dense magnetised plasma structures were driven as to excite the kink mode. After some time in the simulation, the wave energy of the global mode can be seen to be concentrated towards the boundaries of the structures (Figure~\ref{fig:pascoe_2012} right panel), with concentrations close to the locations where $\omega_A({x},y)=\omega_k$. The details of the dual process are similar at a plasma interface \citep{Lee_1986,HOLYAN1988}, which is not so surprising given the fact that the kink mode can be described as a surface Alfv\'en wave. This would suggest that resonant absorption and phase mixing are key features of Alfv\'enic wave damping throughout the solar atmosphere, even if coronal structures are collections of magnetic fields associated with diffuse and complex density structures \citep[as suggested by, e.g.,][]{magyar2016, 2022ApJ...927....1M}.

In terms of contributing to the coronal heating it appears that the actual heating rates, due to the combination of the resonant absorption of kink modes and subsequent phase mixing of rotational motions, are inadequate to compensate for the radiative losses \citep{Pagano_2017,Howson_2020}, even when a realistic broadband driver is employed \citep{Pagano_2019,Pagano_2020}. Further, observational estimates of the frequency-dependent damping rate of propagating kink waves in the quiet Sun suggest that resonant wave damping is likely small \citep{tiwari_2021, Morton_2021}. In coronal holes the Alfv\'enic waves are found to largely undamped \citep[e.g.,][]{Dolla_2008, Banerjee_2009, MORetal2015}. This pattern of weak damping and no damping could reflect that the density differences across the inhomogeneities are small in the quiet Sun and smaller in coronal holes. Further, it would appear that the transfer of energy to the rotational modes is small and any subsequent phase mixing is also weak. 

{We note that there has been some evidence for wave damping in coronal holes above 1.2~$R_{\odot}$ \citep{BEMABB2012,HAHetal2012,Hara_2019}, inferred from a decrease in the non-thermal widths of spectral lines (which is often taken as an indicator of Alfv\'enic waves). The suggested rapid damping is somewhat surprising and there is difficulty in explaining it through known wave damping mechanisms. The results have only been found in Hinode EIS data so far, with the suggestion that uncertainties in scattered light may account for the observed decrease in line widths \citep{Zhu_2021}. Further, \cite{Gilly_2020} suggest that failure to take into account non-equilibrium ionisation and line broadening from the solar wind can lead to misleading inferences from spectral line widths. There have also been other suggestions for the observed decrease in line width \citep[see, e.g.,][]{Cranmer_2018}.}

While the situation does not appear to be favourable for wave heating via phase mixing, the effects pertinent to the stability of the resonance layer can influence energy transfer to small scales.

\subsection{Instabilities}\label{sec:sub_insta}
The role of wave-driven instabilities has received significant focus in the last few years and appears to be an important ingredient for effective wave energy dissipation in the solar atmosphere. The instabilities are localised processes that are entangled with the dynamics and inhomogeneity of the plasma. While there are many instabilities that can occur in a plasma, there has been a particular focus on the Kelvin-Helmholtz (KH) instability in conjunction with propagating Alfv\'enic waves in a cylindrical waveguide.

The KH instability is a fluid instability formed by two fluids undergoing differential shearing motion across an interface, with the instability destroying the shear layer through the development of vortices. The KH instability can also occur in magnetized plasma if the velocity is perpendicular to the magnetic field, although its properties are modified by the magnetic field due to the \textit{frozen-in} condition. However, for a velocity shear parallel to the magnetic field, the instability cannot develop as the magnetic tension has a stabilizing effect \citep{Chan_1961}.  If the magnetic field is sheared across the instability, this can have a stabilizing or destabilizing effect on the KH mode, depending on the properties of magnetic shear \citep{Ofman_1991}. KH has been studied theoretically in the context of coronal plasmas, where it is believed to play an important role in the transition to turbulence and heating \citep[e.g.,][]{HEYPRI1983, Ofman_1994, Karpen_1994}.

\begin{figure}[!t]
    \centering
    \includegraphics[scale=0.45]{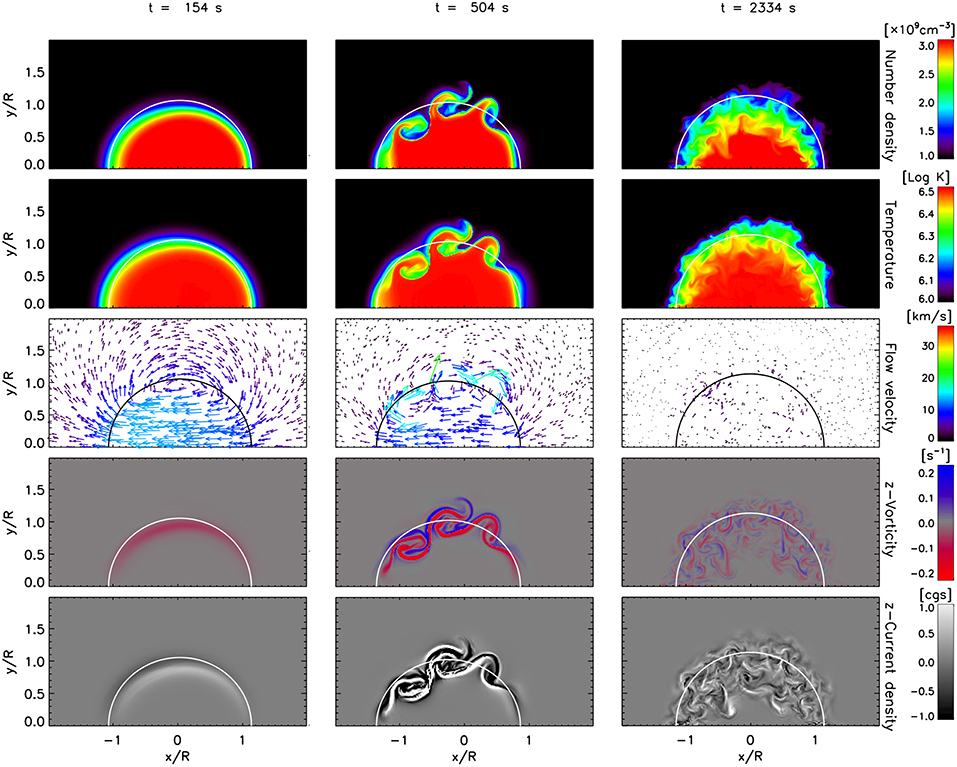}
    \caption{Development of the Kelvin Helmholtz instability at the boundary of a magnetised plasma cylinder. In descending order, the rows show the number density, temperature, flow velocity vectors, parallel vorticity and parallel current. Towards the end of the simulation, the small-scales present are much better observed in the parallel vorticity and current. Reproduced from \citet{antolin2019}.}
    \label{fig:KH_ant}
\end{figure}

With respect to wave motions, it was suggested that because Alfv\'en waves can oscillate on neighbouring magnetic surfaces without interacting, hence can have an arbitrary phase with respect to each other. This would result in shear flows between the surfaces, which could be susceptible to the KH instability. The instability can then enhance the efficiency of the wave dissipation by more rapidly generating small-scales than phase mixing alone \citep{HEYPRI1983,Holl1984, HOLYAN1988}. The development of the KH instability was shown to be present in numerical non-linear studies of the resonant absorption of Alfv\'en waves in inhomogeneous plasma \citep[e.g.,][]{Ofman_1994,Karpen_1994}.

More recently, the development of KH was also shown to be present at the boundary of cylindrical flux tubes undergoing kink motions \citep[an example of this is shown in Figure~\ref{fig:KH_ant}, but see also, e.g.,][]{TERetal2008b,soler_2010,antolin2014,magyar2016,Antolin_2017, karampelas_2019}. The transverse motions of the flux tube through an ambient plasma naturally leads to a shear layer at the tube boundary, which is at an angle to the magnetic field (see the flow velocity in Figure~\ref{fig:KH_ant}). \citet{soler_2010} demonstrated that for a non-twisted cylindrical flux tube with a discrete change in density at the boundary, the dispersion relation (in the thin tube approximation) is given by
\begin{equation}
   \omega \approx \frac{\rho_i}{\rho_i+\rho_e} m\delta v \pm \left[ \frac{(\rho_i v^2_{Ai} + \rho_e v^2_{Ae})}{(\rho_i + \rho_e)} k^2_z - \frac{\rho_i \rho_e}{(\rho_i + \rho_e)^2} m^2\delta v^2\right]^{1/2},
\end{equation}
where $\delta v$ is the velocity jump at the shear layer. In the absence of magnetic twist, the shear flow is perpendicular to the magnetic field. The two terms within the square brackets represent the suppression term (which is essentially the square of the kink speed) and the driving term, respectively. The stability of the shear layer depends upon relative sizes of these two terms. A similar term can be obtained for surface Alfv\'en waves \citep{Hillier_2018}.

As discussed, the resonant conversion of the kink mode to local Alfv\'en modes in the inhomogeneous layer also leads to phase mixing and further increases the shear velocity \citep{antolin2014,antolin2019}. The KH instability is found to develop more slowly in the presence of an inhomogeneous layer \citep{TERetal2008b}. It appears the reason for this is not yet clear (P. Antolin, private communication), but simulations show that the onset times are longer for increasing inhomogeneity length-scales. 
\citet{antolin2019} suggest this is due to the fact that the larger inhomogeneous length scales preferentially excite unstable modes with larger wavelengths, which have slower growth rates. Further, the KH instability distorts the inhomogeneous layer, thereby widening it and enabling the Alfv\'enic motions to become present in a greater portion of the loop \citep{antolin2014}. This process could be severe, and result in complete distortion of an initial circular cross-section \citep[as discussed in Section~\ref{sec:fs_struc}][]{magyar2016}. The development of the KH instability can be delayed by the presence of magnetic twist \citep{soler_2010,Terradas_2018b} or large values of dissipative coefficients \citep{Howson_2017}, but the boundary
of the oscillating loop is always unstable \citep{Barbulescu_2019}.

The presence of the KH instability is also prominent when a cylindrical flux tube is perturbed as to only excite torsional Alfv\'en modes \citep{Guo_2019, D_az_Su_rez_2021}. The development of the instability occurs in much the same manner as the kink mode, suggesting the initial shear from the swaying motion of the kink mode is not a necessary condition to generate the instability in the wave guide. For the KH instability from torsional modes, it is the phase mixing of the Alfv\'en waves that leads to the velocity shears, as suggested
by \citet{HEYPRI1983}. \cite{D_az_Su_rez_2021} undertook a parameter study showing how the onset of the KH instability is influenced by a number of factors, highlighting the size of the inhomogeneous layer across the flux tube and the density contrast can play key roles. Here, the authors suggest the onset of KH instability is delayed for wider inhomogeneous layers as the phase mixing takes longer to develop and, hence, so does the build-up of shear velocities. For density contrast, the development of the instability occurs faster when the density contrast is larger (although after a certain value, the rate of phase mixing is insensitive to the density contrast).

The development of the KH instability appears to be able to provide a substantial heating of the plasma \citep{Karampelas_2017}, unlike the case of resonant absorption and phase mixing. The KH instability can increase the rate at which small-scales are developed from phase mixing alone, hence expediting the transfer of energy to dissipative scales. The development of the KH instability and subsequent heating can be increased when multiple modes are initially present \citep{Guo_2019}. It was shown by \citet{Shi_2021} that the development of KH instability in simulations of kink oscillations can provide heating rates, that under specific conditions are sufficient enough to balance radiative losses in the quiet Sun. 

It should be noted that the majority of these studies focus on standing kink modes, which have, to date, only been identified in active regions. In contrast, observations of Alfv\'enic waves in the quiet Sun and coronal holes appear to show only propagating modes \citep{MORetal2015,tiwari_2021,Morton_2021}. It would appear that the KH instability is also able to develop in numerical simulations of a system containing only propagating Alfv\'enic waves, although the low resolution of the simulations prevented detailed study of the instability \citep{Pagano_2019}. However, \citet{Zaqarashvili_2015} suggest propagating waves are stable to the KH instability as the magnetic and velocity perturbations of Alfv\'enic modes are in anti-phase, meaning that that magnetic field stabilises the system. Hence, there is still some ambiguity around what conditions are required for the KH instability to develop in a system primarily containing propagating waves.


\subsection{Turbulence}\label{sec:sub_turb}
Turbulent cascades have long been considered a promising candidate for dissipation of Alfv\'en waves to heat coronal loops \citep[e.g.,][]{1983NASCP.2280...5H,2007ApJ...657L..47R,VANBALLetal2011,Verdini_2012} and accelerate the solar wind \citep[e.g.,][]{1986JGR....91.4111H,1989PhRvL..63.1807V,VERDetal2010}. Inhomogeneity is essential in many models of Alfv\'enic wave propagation from the corona to the solar wind. However, it is predominately assumed to be along the field only. More recently the influence of perpendicular inhomogeneity has become of interest. So, to set the scene for this contemporary take, we briefly discuss salient aspects of the long-studied Alfv\'en wave turbulence \citep[the work in this area is extensive and spans many decades, and an excellent review can be found in][]{BRUCAR2005}.

The incompressible MHD equations that describe the evolution of the magnetic field and the velocity fluctuations can be written in terms of the so-called Els\"asser variables, $\vec{z}^{~\pm}=\vec{v} \mp \vec{v_A}$ \citep{Elsasser_1950}, where $\vec{v_a}=\vec{B}/\sqrt{\mu_0\rho}$. It can be shown that
they can be reduced to the following equation:
\begin{equation}
\frac{\partial \vec{z}^{~\pm}}{\partial t} + (\vec{z}^{~\mp}\cdot\nabla )\vec{z}^{~\pm}=-\frac{1}{\rho}\nabla P- \vec{v_A}(\nabla\cdot \vec{v_A}),
\label{eq:turb}
\end{equation}
where $P=p+\frac{B^2}{2\mu_0}$ is the total pressure. The second term on the left hand side describes 
the non-linear interaction of counter propagating waves along the mean magnetic field, and is the 
key factor in enabling a turbulent cascade. The non-linear interaction  deforms the counter-propagating 
wave packets and generates the small-scale structures. A typical assumption for Alfv\'enic turbulence in the corona and solar wind is that there is strong mean magnetic field, such that the amplitudes of perturbations are small in comparison to the background magnetic field \citep[referred to as Reduced MHD]{1976PhFl...19..134S}. With such a strong anisotropy, the cascade is primarily in the perpendicular wavenumbers with parallel cascades suppressed, compared to, e.g., a case with a weaker background field \citep{2003LNP...614...28O}. 

Typically, the inhomogeneity is only assumed to be along the magnetic field (due to gravitational stratification), and the Alfv\'enic waves are subject to non-WKB reflection due to the gradient in the Alfv\'en speed \citep{HEIOLB1980,1993A&A...270..304V}. In the presence of density inhomogeneities along the magnetic field, the final term on the right-hand side of Eq.~\ref{eq:turb} leads to the reflection of Alfv\'en waves\footnote{Perhaps unsurprisingly, there is only reflection in an incompressible plasma if there is a gradient density. One may have thought this should
be the case for variations in Alfv\'en speed due to the magnetic field gradients, however this is not the case \citep{Magyar_2019}.}\citep{HEIOLB1980,1993A&A...270..304V}. This term is vital for the development of turbulence in the solar wind, given it is expected that Alfv\'enic waves are launched predominantly at the Sun. Without this reflection, there would be no counter-propagating waves to interact. Much of the solar-focused literature on Alfv\'enic wave turbulence generally only includes this form of reflection to drive the turbulence, with some compressible simulation also able to include Alfv\'en wave reflection to due to parametric decay instability \citep[e.g.,][]{1969npt..book.....S,Goldstein_1978,Shoda_2018,Shoda_2018b,R_ville_2018}.

\medskip

A compelling development in the discussion of Alfv\'enic wave turbulence is {recent work that has included the presence of perpendicular density inhomogeneities across the magnetic field. Largely this has been neglected in turbulence studies.} However, the presence of the inhomogeneity enables a self-cascade of the Alfv\'enic waves, given the name `uni-turbulence' \citep{Magyar_2017,Magyar_2019}. As discussed in Section~\ref{sec:wt_in_perp}, the presence of inhomogeneities perpendicular to the magnetic field leads to the existence of surface Alfv\'en modes in the presence of a discontinuity, or a quasi/global Alfv\'enic waves in the presence of a continuous variation \citep[e.g.,][]{Goossens_2002,GOOetal2012}. \citet{Magyar_2017} showed via numerical simulations of unidirectional Alfv\'enic modes that both Els\"asser components are present, propagating in the same direction. This leads to the cascade of wave energy in the perpendicular direction, with the timescale for the cascade dependent on the size of the density contrast \citep{Van_Doorsselaere_2020}, namely:
\begin{equation}
    \tau = \sqrt{5\pi}\frac{1}{\omega a}\frac{\rho_i+\rho_e}{\rho_i-\rho_e},
    \label{eq:uni_turb}
\end{equation}
where $a$ is the normalised oscillation amplitude. The presence of both components is due to the fact that, as emphasised already, in inhomogeneous 
media 
MHD waves have mixed properties. \citet{2019ApJ...873...56M} demonstrate that when compressibility and inhomogeneity of the plasma are taken into account, the Els\"asser formalism cannot be used to separate parallel and anti-parallel propagating waves. In a homogeneous, compressible plasma the slow and fast modes are necessarily described by both components. Hence, waves with mixed properties should also have to be described using both Els\"asser components. We refer readers to \citet{Magyar_2019} for a discussion of the subtleties of this phenomena applied to Alfv\'enic modes. 

The study of this turbulence phenomenon and its role in plasma heating is still in its infancy but it would appear to be an exciting avenue to aid in the transfer of wave energy to dissipation scales. Application to the damping of standing kink modes in active regions suggest that it could explain an observed non-linear relationship between wave amplitude and damping times \citep{GOONAK2016, Van_Doorsselaere_2021}. This phenomenon could also be a dominant source of energy transfer in open field regions, i.e., coronal holes, where the Alfv\'enic wave propagation appears pre-dominantly uni-directional \citep[i.e., anti-sunward][]{MORetal2015}. For closed, loop-like magnetic fields, observations suggest that counter-propagating waves are the norm \citep{tiwari_2021}. These waves are likely driven at both foot-points of the loops (by granulation or mode conversion) and suffer strong wave reflection at the transition region; ensuring a constant supply of counter-propagating waves. Hence, uni-turbulence would play a complementary role. However, \citet{MORetal2021} examined a number of loops in the quiescent corona and inferred that the density difference across the inhomogeneities must be small to explain the apparently weak frequency-dependent wave damping of the observed kink modes (i.e., due to resonant absorption). This would also imply that time-scales for energy transfer by uni-turbulence may also be long given Eq.~\ref{eq:uni_turb}.  

\section{Waves in the chromosphere}\label{sec:chrom}

In order to reach the corona, the waves originating in the photosphere, must pass through the gauntlet of the chromosphere. This layer remains one of the least understood regions of the Sun's atmosphere due to many complex physical phenomena that arise there. However, there is a general agreement that the chromosphere is highly dynamic and continuously exchanging mass and energy with the corona
\citep[for reviews see, e.g.,][]{2006ASPC..354..259J,2019ARA&A..57..189C}. Moreover, this complex region is populated by fine-scale structures \citep[see review by][]{TSIetal2012}, that bridge the photosphere with the corona and act as conduits for plasma and wave energy transfer across the transition region. A peculiar aspect of this layer is associated with moderate rise in its thermal profile (Fig. \ref{fig:T_rho_atmosphere}), as compared to the photosphere. This maintains temperatures of the order of $\sim$10$^{4}$~K in the chromosphere, which enables much of the hydrogen and helium to remain neutral up until the transition region (Figure~\ref{fig:ion_neutral}). In these conditions, partial ionisation of plasma becomes a critical factor that can have a strong effect on the dynamics \citep[see reviews by, e.g.,][]{Leake_2014, Mart_nez_Sykora_2015}. In combination with large gradients along the magnetic field, it indicates that most of the non-thermal energy remains in the chromosphere.

The observed dynamics of chromospheric features reflect the presence of confined MHD waves that emanate from the photosphere. Alfv\'enic waves are routinely observed and are often classified as a particular wave mode depending upon their observational characteristics \citep[see review by][]{Verth_2016}. They carry mass, energy and momentum across this layer. However, in the lower solar atmosphere there are two major factors that influence Alfv\'enic wave propagation, Namely the presence of partial ionisation (which leads to wave damping) and the gradient of the Alfv\'en speed (Fig.~\ref{fig:mag_field}, leading to reflection and non-linear effects).

\subsection{Partial ionisation}\label{sec:sub_PI}

\begin{figure}[!ht]
    \centering
    \includegraphics[scale=0.6]{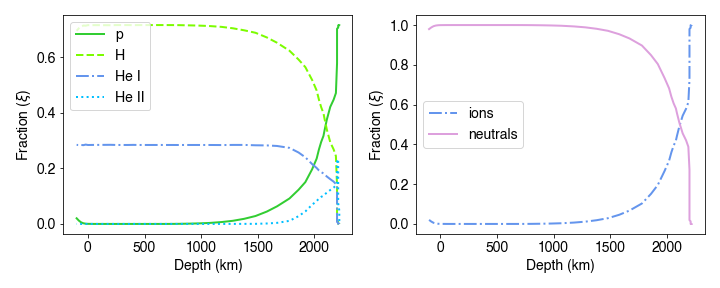}
    \includegraphics[scale=0.6]{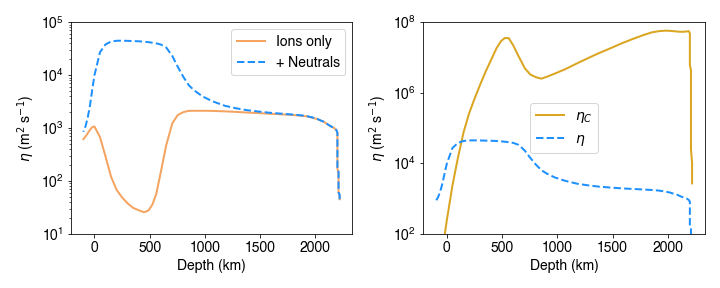}
    \caption{Properties of the partially ionised solar atmosphere based on the FAL-C atmosphere. The top left panel displays the relative fraction of different components of the plasma. The top right plasma composition in terms of the relative fractions of ions and neutrals fractions. The bottom row shows the resistivity through the lower solar atmosphere. The left panel displays the classical electrical resistivity (i.e., electron collisions). The right panel shows the values of resistivity for electrical (including neutrals) in blue dash and Cowling resistivity in solid orange.  }
    \label{fig:ion_neutral}
\end{figure}

The process of ionisation primarily depends on the (inelastic) collisions between the different constituent particles of the plasma. In Figure~\ref{fig:ion_neutral}, we illustrate the ionisation degree for the dominant components required for a multi-fluid solar plasma (i.e., hydrogen and helium), along with the fractions of ions and neutrals. The main influence that partial ionisation has on wave propagation arises through the generalised Ohm's Law, namely;
\begin{equation}
    \frac{\partial \vec{B}}{\partial t} = -\nabla\times(-\vec{v}\times\vec{B}
    +\eta j_{\parallel}+\eta_p j_{\perp}+\frac{1}{en_e}\vec{j}\times\vec{B})
    \label{eq:gen_ohms}
\end{equation}
where, $j_{\parallel}$ and $j_{\perp}$ are the parallel and perpendicular current respectively. All other symbols have the usual meaning. The second term on the right is the Joule heating (with $\eta$ the electrical resistivity) that describes the electron collisions with ions and neutrals. The third term is due to the additional collisions of ions and neutrals  (and $\eta_p$ is the Pedersen resistivity), while the fourth term is the Hall term. The Pedersen resistivity can be decomposed into the perpendicular components of the electrical resistivity and the Cowling resistivity\footnote{Also referred to as the coefficient of ambipolar diffusion, $\eta_A$.} , i.e. $\eta_p=\eta+\eta_C$. The Cowling resistivity for a Hydrogen-Helium plasma is given by \cite[e.g.][]{Zaqarashvili_2013},
$$
\eta_C = \frac{B^2}{\mu_0}\frac{\alpha_{He}\xi_H^2+\alpha_{H}\xi_{He}^2+\alpha_{HeH}(\xi_H+\xi_{He})^2}
{(\alpha_{H}\alpha_{He}+\alpha_{H}\alpha_{HeH}+\alpha_{He}\alpha_{HeH})},
$$
where, $\xi_n$ is the fraction of species $n$, and $\alpha_{n_1n_2}$ is the coefficient of friction between different species $n_1$ and $n_2$ (note that a subscript of a single species means the sum of coefficients for ion-neutral collisions associated with the given species).

Studies to evaluate the impact of ion-neutral effects in the solar atmosphere were pioneered by \citet{1956MNRAS.116..314P} and \citet{OST1961}, where these authors first highlighted their importance as a wave damping mechanisms in the chromosphere. Decades later, their initial research was further extended  for the analysis of Alfv\'en wave propagation in a partially ionised single-fluid (i.e., hydrogen) MHD model \citep{1998A&A...338..729D, Leake_2005}, incorporating more realistic estimates for plasma parameters. It was suggested that wave damping due to Ohmic diffusion is dominant over ion-neutral damping in low photosphere and chromosphere, but can be neglected due to the long damping lengths. However, in the upper chromosphere, the damping is sensitive to ion-neutral effects (which can be inferred the magnitude of the Cowling resistivity in Figure~\ref{fig:ion_neutral}) and is effective for high-frequency waves ($f > 0.01$~Hz). However, low frequency waves remain unaffected by the presence of neutrals and experience no damping due to this mechanism. Further, \citet{Leake_2005} also demonstrated that increasing the magnetic field strength can weaken the damping of waves due to ion-neutral effects. Further, weak damping of Alfv\'enic waves in partially ionised atmosphere can also create drag forces that can lift the cold/dense plasma against the gravity and generate spicule-like structures in the vertical direction 
\citep{Haerendel_1992,1998A&A...338..729D,James_2003}.

Naturally, one should also expect to have additional viscosity effects due to ion-neutral interactions, arising through the momentum equation. \citet{2004A&A...422.1073K} provided a quantitative comparison of the relative magnitudes of damping due to viscous and collisional terms, based on the analytic expressions for the wave damping times. It was demonstrated that ion-neutral effects are more efficient compared to viscous forces for Alfv\'en waves. \citet{Soler_2015} extended this line of work by incorporating multiple potential damping mechanisms into an analytic investigation, and confirmed that viscosity plays no important role in wave damping throughout the chromosphere. Using a single fluid description, \citet{Soler_2015} included the effects of partial ionisation involving both neutral hydrogen and helium. For Alfv\'en waves, they showed that Ohmic diffusion dominated the damping at low heights, while ambipolar diffusion became more important at greater heights in the chromosphere. This result was in qualitative agreement with \citet{1998A&A...338..729D} and \citet{Leake_2005}, although \citet{Soler_2015} suggested that Ohmic diffusion is actually an important damping mechanism in the low chromosphere \citep[see also][]{Goodman_2011,Tu_2013,2016ApJ...817...94A}. The cause of this difference is mainly due to taking into account the role of electron-neutral collisions when calculating the Ohmic resistivity, which leads to a significant increase in the magnitude of the coefficient (bottom left panel, Figure~\ref{fig:ion_neutral}). \citet{Soler_2015} also highlighted the role of the magnetic field strength, which determines at which heights ambipolar damping overtakes Ohmic damping. 

\medskip 

Although insightful, the use of single fluid approximations to understand the wave behaviour in the photosphere and chromosphere has been shown to be incorrect. This aspect was examined in  detail by \citet{ZAQetal2011}, who investigated MHD wave propagation in a two fluid model consisting of neutral hydrogen and ions. They showed that for time-scales less than the ion-neutral collision time, both fluid species tend to behave independently and a single-fluid approximation does not remain valid any more. Importantly, the damping of high frequency waves is a lot weaker than suggested by the single fluid approach. Their results again confirmed that low-frequency waves were not strongly influenced by partial ionisation of the plasma. More generally, \citet{SOLetal2013} showed that the damping of waves is most efficient when the wave frequency is comparable to the ion-neutral frequency.

Given the sizable fraction of neutral helium in the lower solar atmosphere, \citet{2011A&A...534A..93Z, Zaqarashvili_2013}, extended the analysis with the inclusion of three fluids. The addition of helium species was found to have an additional impact on the damping of high frequency Alfv\'en waves. However, this mechanism is strongly dependent on the plasma temperature, with larger wave damping rates at higher temperatures ($1-4\times 10^4$~K), when ionisation of hydrogen becomes prevalent (see Figure~\ref{fig:ion_neutral}). Further, \citet{Soler_2015b} found that there exists a particular wavelength range for which Alfv\'en waves are overdamped (10$\lesssim \lambda \lesssim$1000~m), resulting in localised energy dissipation and thus restricting the propagation as Alfv\'en wave. 

\medskip

In an effort to incorporate perpendicular inhomogenities into studies of wave propagation in a partially ionised plasma, \citet{SOLetal2012} investigate the influence of ion-neutral collisions on kink wave propagation in a cylindrical flux tube in a two-fluid study. The influence of the ion-neutral interaction was incorporated in the momentum equation through frictional terms. With this formalism, \citet{SOLetal2012} found that ion-neutral collisions appear to have little impact on kink wave propagation and resonant absorption remained the dominant damping mechanism.  Interestingly, they found that the damping lengths due to resonant absorption, for kink waves with periods less than 10~s, are smaller than or of the same order as the height of the chromosphere. Hence, they suggested that only waves with periods greater than 10~s are able to reach the coronal level in the form of kink motions, while the shorter period waves would reach the corona as rotational motions.

\medskip

Finally, we note that use of the multi-fluid approach has been suggested to be invalid for short wavelengths. This was first highlighted by \citet{Soler_2015b} who demonstrated that the multi-fluid approach, which essentially treats each plasma species separately, is not applicable to high-frequency waves below of heights $\sim900$~km. They suggest a kinetic description of the plasma is required as both ions and electrons collide too frequently with the neutrals, meaning that the thermal properties of each species are not independent.

\begin{figure}[!ht]
    \centering
    \includegraphics[scale=0.6]{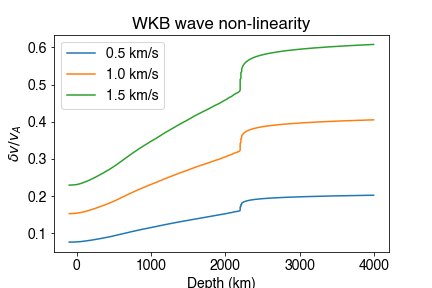}
    \caption{The ratio of wave velocity amplitude to Alfv\'en speed is shown as a function of height. The curves are calculated using a WKB approximation for wave amplification in an ideal plasma {using the density from the FAL model and the magnetic field from the empirical model, see Figure~\ref{fig:T_rho_atmosphere}.}}
    \label{fig:non_linear}
\end{figure}

\subsection{Alfven speed gradient}
Aside from the longitudinal homogeneity in ionisation state, there is also a longitudinal inhomogeneity in the Alfv\'en speed through the chromosphere due to the decrease in density (Figure~\ref{fig:T_rho_atmosphere}). This variation leads to amplification of the waves propagating from the photosphere to the corona. The relationship between wave amplitude and density can be obtained under a WKB approximation, namely $\mid\delta v\mid\propto \rho^{-1/4}$. Hence, the upwardly propagating Aflv\'en waves can become non-linear, demonstrated by the increase of the ratio of the wave amplitude to the Alfv\'en speed ($\mid\delta v\mid/v_A$) with height\footnote{We note the form of this profile is dependent upon the functional form of the magnetic field used and can decrease in the corona \citep[c.f. Figure 4 in][]{Wang_2020}.} (shown in Figure~\ref{fig:non_linear}). Given that the profiles in Figure~\ref{fig:non_linear} are found using the WKB approximation, they are not valid for waves with wavelengths comparable to {or greater than} the density scale height in the lower solar atmosphere. Moreover, the calculation of the profiles neglects any physics that would act to damp these waves, e.g. ion-neutral effects, Ohmic diffusivity, and mode conversion (of which we now discuss). Hence, in reality the increase in wave amplitude is less rapid than shown in the figure.

The large amplitude of the Alfv\'en waves leads to an increase in the ponderomotive force, which is able drive motions parallel to the magnetic field. This enables the mode conversion of Alfv\'enic modes to both fast and slow magnetoacoustic modes \citep[e.g.,][]{HOLetal1982, Hollweg_1992}. Both the fast and slow waves also develop into shocks in the upper chromosphere and low corona. When the magneto-acoustic shock occurs in the chromosphere, there is an upward displacement of the chromosphere and transition region creating spicule-like structures \citep[e.g.,][]{HOLetal1982, Hollweg_1992}. The slow shocks driven by Alv\'en waves are found to be very effective at heating the lower solar atmosphere \citep{Matsumoto_2012,2016ApJ...817...94A,Wang_2020}.  \citet{2016ApJ...817...94A} suggest that the relative heating rates due to shock heating are four orders of magnitude greater than the Pederson resistivity. However, the heating rate in their 1.5D simulations may be enhanced by shock coalescence. The fast and slow mode waves produced by Alfv\'en waves are also capable of heating the corona \citep{Hollweg_1992,Kudoh_1999,Moriyasu_2004,Antolin_2008,ANTSHI2010}.

\begin{figure}
    \centering
    \includegraphics[trim=0 -0.5cm 0 0,clip,scale=0.6]{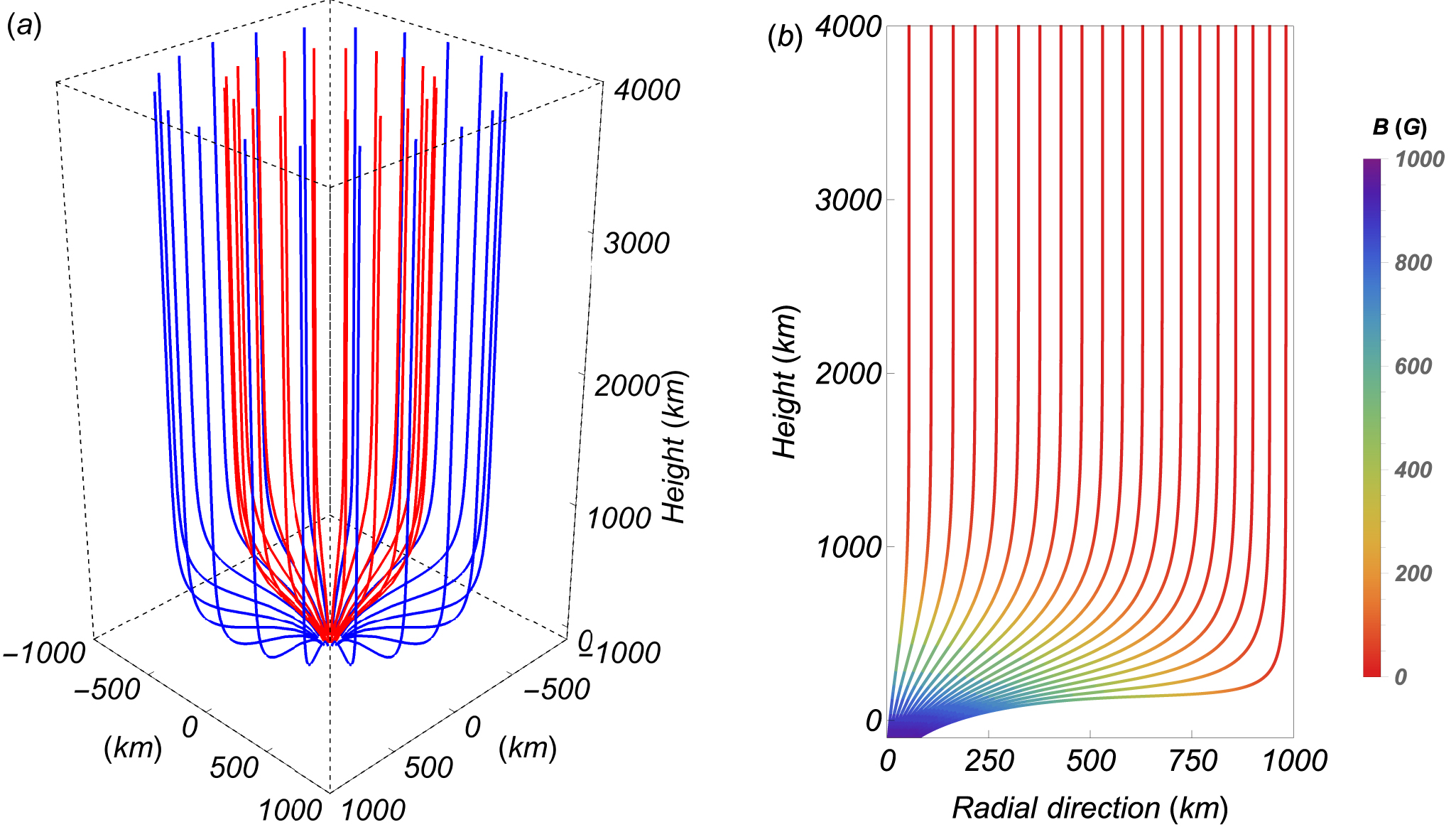}
    \includegraphics[trim=0 0 0 5.6cm,clip,scale=0.6]{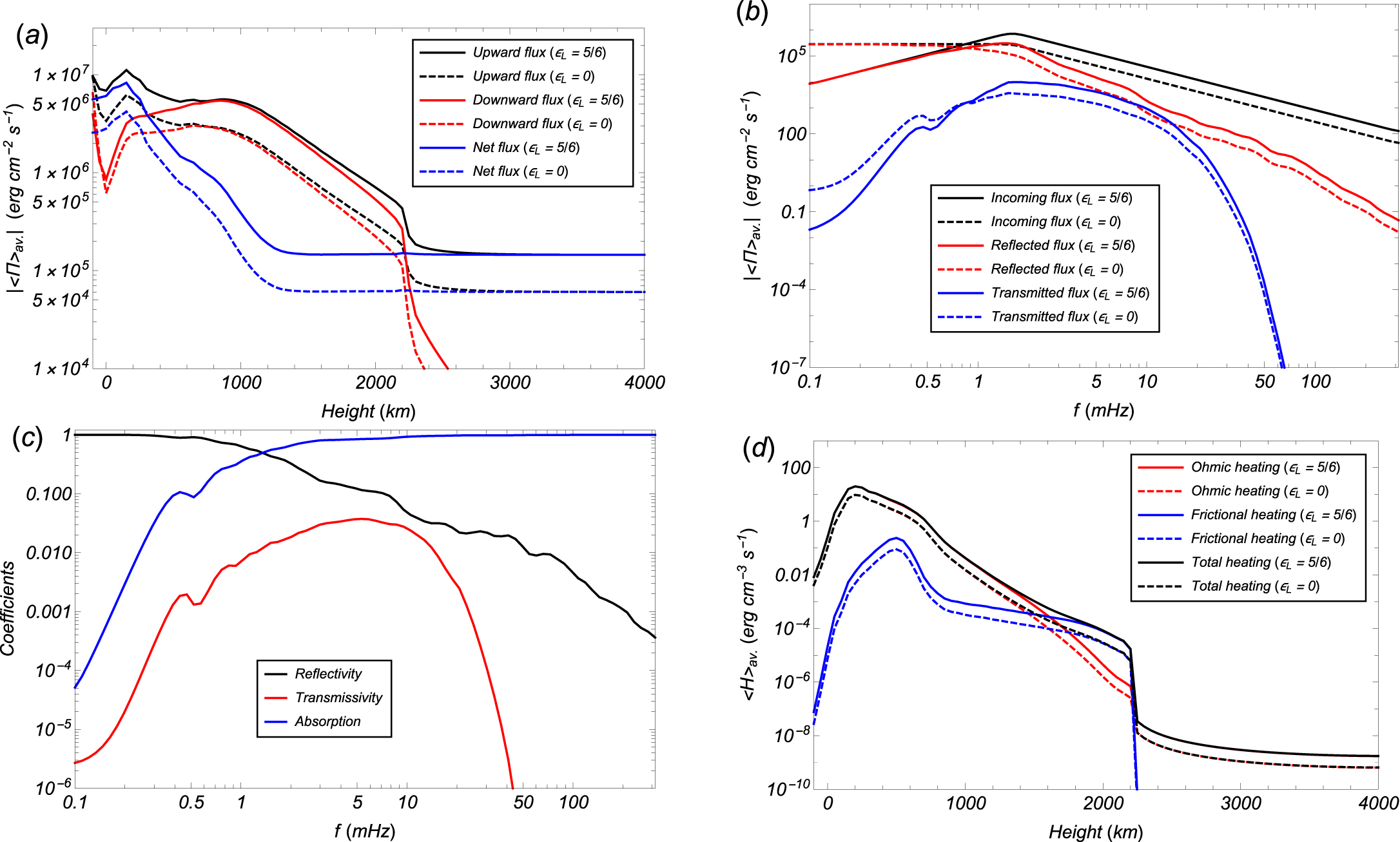}
    \caption{The structure of the magnetic field from the model of torsional Alfv\'en wave propagation through the lower solar atmosphere from \citet{2019ApJ...871....3S}. Panel a shows the expanding magnetic field in 3D. Panel b displays a cross-sectional profile of the magnetic field in the radial direction. Panel c is the horizontally averaged profiles for reflection, transmission and absorption as a function of wave frequency. Panel d shows the contribution to the wave heating rates as a function of height in the model atmosphere. Figures reproduced from \citet{2019ApJ...871....3S} with permission of the AAS.}
    \label{fig:soler_figs}
\end{figure}

\medskip
The variation in Alfv\'en speed also leads to significant reflection of the waves in the chromosphere, and especially at the transition region. The reflection of waves occurs if the wavelength of the wave is comparable or greater than the length-scale over which there is appreciable variation in Alfv\'en speed \citep[e.g.,][]{Ferraro_1958,1993A&A...270..304V}. Hence there is frequency filtering effect, with low frequency Alfv\'en waves affected to a greater extent than higher frequency waves \citep[see Figure~\ref{fig:soler_figs} panel c for a curve of reflection versus frequency; details of calculation given in][ ]{2017ApJ...840...20S,2019ApJ...871....3S}. The wave reflection also sets up a system of counter propagating waves in the chromosphere, which, as discussed in Section~\ref{sec:sub_turb}, enables Alfv\'enic wave turbulence to develop \citep[e.g.,][]{VANBALLetal2011}.

In general, the transition region provides a substantial barrier to all waves, significantly influencing the transmission to the corona \citep{Hollweg_1981,Verdini_2012}, with \citet{CRAVAN2005} suggesting the reflection rate is on the order of 95\%. {Fully dynamical simulations suggest the wave reflection might be lower, but still substantial, with 85\% of the initial wave energy flux reaching the corona \citep{SUZINU2005,Suzuki_2006}. In the \cite{SUZINU2005,Suzuki_2006} model, the transition region is dynamic due to intermittent shock heating in the chromosphere and the sharp density gradient is smoothed out.} \citet{2016ApJ...817...94A} estimate of the efficiency of wave transmission to the corona by comparing the Poynting flux at the photosphere against the Poynting flux entering the corona. In their incompressible simulation, the transmission to the corona is a relatively smooth function of frequency. The transmission is most effective for intermediate frequencies, peaking at around 40\% of the initial wave flux for frequencies of 0.1-1~Hz, with the reduction in transmission for higher frequencies related to wave damping. The degree of magnetic expansion can also influence the transmission curve, with larger rates of expansion shifting the transmission peak to lower frequencies \citep{Similon_1992,2017ApJ...840...20S}. In Figure~\ref{fig:soler_figs} we show the reflection, transmission (to corona) and absorption (i.e., dissipation due to wave damping) curves for a model of Alfv\'en wave propagation from \citet{2019ApJ...871....3S} model \citep[which show a qualitatively similar behaviour to that in][]{2016ApJ...817...94A}.

\subsection{Perpendicular inhomogeneity}\label{sec:chrom_perp}
From the previous discussions, it is clear that Alfv\'enic wave propagation is strongly influenced by the perpendicular inhomogeniety of plasma, hence it is worth touching upon its influence in a partially ionised plasma. While there are a number of 2D and 3D models of wave propagation through the lower atmosphere that include or develop some perpendicular inhomogeniety \citep[e.g.,][]{Shelyag_2016,Ku_ma_2020,2021A&A...649A.121B,Breu_2022}, a fascinating insight can be found in \citet{2019ApJ...871....3S}. The model focuses on the propagation of torsional Alfv\'en modes from photosphere to corona but is quite restrictive; it is an incompressible, 2.5D model of linear waves. However, it is an insightful culmination of a series of works that carefully increases the models complexity and isolates the impact of aspects of wave physics \citep[see][]{SOLetal2013,Soler_2015,2017ApJ...840...20S,2019ApJ...871....3S,2021ApJ...909..190S}. In \citet{2019ApJ...871....3S}, the perpendicular inhomogeniety arises through the implementation of a 3D magnetic field, which is shown in Figure~\ref{fig:soler_figs}. Through the chromosphere the magnetic field expands but that expansion rate varies as a function of radius. As such, there is a variation of the Alfv\'en speed in the perpendicular direction, as well as the parallel direction. This leads to phase mixing of the Alfv\'en waves and the development of small-scale shears in the velocity and magnetic perturbations across the flux tube. In turn, this leads to an enhancement of local currents and increases the rate of Ohmic diffusion. 

Some of key results from the \citet{2019ApJ...871....3S} are shown in Figure~\ref{fig:soler_figs}, which also nicely summarises aspects of the previous discussions.  In panel c, the reflection, transmission and absorption (i.e. wave damping) of the waves are shown as a function of frequency. There is clear frequency dependency for all quantities. It can be immediately seen that the transmission of wave energy to the corona is drastically reduced compared to the estimates from the lower dimensional models discussed previously. The peak of the transmission curve is close to 5~mHz, and only $\sim$1\% of the photospheric wave energy flux is able to reach the corona. The reason for the decrease in transmission is due to the increase in Ohmic heating arising from the phase mixing. Ohmic heating now dominates the frictional heating (due to ion-neutral collisions) through the majority of the lower atmosphere (panel d), and it is only surpassed in the upper chromosphere ($\sim1700$~km). In this model, the Ohmic heating rate has increased while the frictional heating rate is similar to previous lower dimensional models \citep{2017ApJ...840...20S}. For an injected wave flux at the photosphere with a realistic total energy of $~10^9$~ erg cm$^{-2}$ s$^{-1}$, the wave flux entering the corona turns out to be only a factor of two smaller than the expected energy flux required to sustain coronal temperatures in the quiet Sun.

We note that the sensitivity of this result to the magnetic geometry, namely the height of the most significant expansion, is not entirely clear but could play a role in determining the relative contributions of the two heating mechanisms. Additionally, given the strong influence of shock heating in the 1.5D models of \citet{2016ApJ...817...94A}, it remains to be seen if the findings from \citet{2019ApJ...871....3S} are still valid for non-linear wave propagation.

\medskip
Another interesting work comes from \citet{Gonz_lez_Morales_2020}, who take a different approach to examining the impact of inhomogeneities on wave propagation. The framework is single fluid 3D MHD but incorporates the
generalised Ohm's Law (Eq.~\ref{eq:gen_ohms}) and additional terms in the energy equations to describe a partially ionised plasma. Instead of focusing on a single flux-tube, a small region of the quiet Sun is modelled and includes solar magneto-convection with a near self-generated magnetic field. This naturally generates magnetic flux tubes through the concentration of flux in the intergranular lanes. The model does not reach the corona but is confined to the lower solar atmosphere. The realistic nature of the magneto-convection means that there are gradients perpendicular to the magnetic field direction throughout the simulation domain.
The paper does not discuss phase mixing of Alfv\'en waves or associated instabilities (although they should be present) but does highlight the important role of two other effects, ambipolar diffusion and the Hall term. In common with \citet{2019ApJ...871....3S}, \citet{Gonz_lez_Morales_2020} find that ion-neutral effects can lead to efficient absorption of Poynting flux. Due to the highly inhomogeneous nature of the plasma, MHD waves and associated instabilities (e.g., Kelvin-Helmholtz) enhance vorticity \citep[e.g.,][]{2019FrASS...6...20G,Howson_2019}, which in turn increases the magnitude of the currents \citep[e.g.,][]{Howson_2021} and increasing the Ohmic and ambipolar heating rates. Futhermore, the inclusion of the Hall term is found to result in a doubling of the
Poynting flux at the upper chromosphere (compared to without its presence), which is partially due to a Hall-induced mode coupling of fast and Alfv\'en waves \citep[see also][]{Cally_2015,Gonz_lez_Morales_2019}.

\section{Summary}

The solar atmosphere is a complex plasma environment that is characterised by the networks of magnetic field that thread it, and the large variations in plasma parameters with height. The temperature, density and magnetic field vary by orders of magnitude from the photosphere to the corona and solar wind, and makes the Sun's atmosphere an interesting laboratory for studying MHD wave propagation. In addition to the large-scale variations, the plasma is also observed to be highly structured on small-scales ($\sim$100's~km) perpendicular to the magnetic field (although there could be much finer structuring below the current limits of instrumentation), likely the result of localised heating events that are believed to take place throughout the atmosphere. As we have hopefully highlighted through the review, the presence of plasma inhomogenities in the solar atmosphere produces a fantastic array of physics that impacts upon wave propagation through the magnetised plasma.

The main motivation for many investigations of Alfv\'enic waves is to understand their role in the heating of the atmospheric plasma and the acceleration of the solar wind \citep[they are also proving useful for magneto-seismology and estimates of the coronal magnetic field, e.g.,][]{Anfinogentov_2019,YANGetal2020}. For the waves to contribute to these processes, they need to propagate upwards from the photosphere into, and through, the partially ionised, magnetically-dominated chromosphere. A fraction of the Alfv\'enic waves are able to cross the barrier provided by the dramatic drop in density at the transition region, reaching the corona and beyond. We have discussed various details of the journey, but typically in isolation. This somewhat reflects how the research has been undertaken to date. As mentioned at the beginning of this review, it is challenging to include all the necessary physics into a single atmospheric model. Hence, to date, many studies have focused on individual problems in isolation in order to make progress (this, of course, is always the way). So to wrap up this review, we highlight a few representative state-of-the-art numerical studies and note the current tensions in the modelling.

\smallskip 

There are many large-scale models that examine wave propagation, focusing on, e.g., wave heating/acceleration of the solar wind over 10's of solar radii, or global simulations of coronal heating. While demonstrating the feasibility of wave-based plasma heating/wind acceleration via turbulence, one of the main sacrifices these models make is that they largely ignore the role of perpendicular inhomogeneity; missing resonances, phase mixing and instabilities. There are also other restrictions required to make these numerical simulations feasible.  
For example, the latest Alfv\'en wave solar wind simulations have been able to incorporate compressibility into a 3D system to examine the non-linear coupling to slow modes, but only starting at the coronal base \citep[e.g.,][]{Shoda_2019} or includes a simple lower atmosphere with a reduced numerical resolution \citep{Matsumoto_2020}. Another examines the propagation of Alfv\'en waves from the lower atmosphere and the onset of parametric instabilities, but using a 1D single fluid approach ignoring partial ionisation \citep[e.g.,][]{R_ville_2018}. Moreover, a global model of coronal heating based on Alf\'ven wave turbulence \citep[the Alfv\'en wave solar model - AW-SoM,][]{van_der_Holst_2014} is able to reproduce global EUV emission reasonably well, but only includes wave propagation from the upper chromosphere and uses Reduced MHD and WKB wave propagation (with correction terms) to predict wave heating rates. Also, smaller-scale Alfv\'enic-wave-driven turbulence models, i.e. focusing on single flux tubes, appear to be able to reproduce necessary heating rates for chromospheric/coronal plasma, but only include a basic photosphere/chromosphere without dynamics or multi-fluid effects \citep[e.g.,][]{Matsumoto_2018}, and are often based on Reduced MHD \citep[e.g.,][]{VANBALLetal2011,Verdini_2012}.

\smallskip 

There are also a wealth of studies, some of which we discussed here, that concentrate on the impact of non-ideal
phenomena and instabilities on wave propagation; requiring high-resolution simulations to capture the relevant scales. Hence the models are typically confined to the lower atmosphere or are single, highly idealised flux tubes, with many of the latter models confined to the coronal section or include only a basic lower atmosphere (i.e., no dynamics or partial ionisation). For example, as discussed in Section~\ref{sec:chrom_perp}, \citet{Gonz_lez_Morales_2020} undertook hydrodynamic convection driven 3D simulations of lower atmosphere including a generalised Ohm's law and examining the impact of ambipolar diffusion and the Hall term on Alfv\'enic waves. However, the simulation domain was confined to heights below 1.5~Mm. To include a corona, \cite{2019ApJ...871....3S} model a single flux tube with a temporally stationary atmosphere based on the FAL atmosphere, solving linearised multi-fluid wave equations. The linearisation obviously means that shocks, mode conversion and momentum transfer are not present, all of which are representative of the observed chromosphere. Moreover, most models with a multi-fluid chromosphere do not include time-dependent hydrogen ionisation, which acts over relatively long time-scales \citep[e.g.,][]{LEEetal2007}. Hence the chromospheric ionisation fraction and temperature are different from equilibrium which would influence the efficacy of various multi-fluid phenomena, e.g., Cowling resistivity.

\smallskip

The final note to end on is a comment on the observations of Alfv\'enic waves in the lower solar atmosphere and corona. To date, the observational studies have been relatively few (excluding work on the standing kink modes in active regions), and little is known from an observational perspective. Over the last three decades there have been a number of indirect measurements of their presence through spectral line broadening \citep[e.g.,][]{BANetal1998,Banerjee_2009,HAHetal2012} in coronal holes and the quiet Sun corona. It took the advent of high-resolution observations to measure the Alfv\'enic
waves directly \citep{TOMetal2007,DEPetal2007,JESetal2009,MCIetal2011,HILetal2013,MORetal2012c}, but the number of studies focused on analysing the details of propagating waves is few. Those that have been performed have largely focused on measuring the typical properties (e.g., amplitude, period) in a single region and providing energy estimates of the waves in the structures examined (e.g., coronal loops, spicules). However, the details of excitation, evolution and dissipation are scant. Part of this issue is due to the challenge of connecting the observed dynamics through the different regions of the atmosphere. Some of the issue lies in observing the chromosphere, which is complicated by extended formation heights of the resonance lines \citep[requiring narrow spectral filters to isolate different heights, e.g.,][]{JUDCAR2010} and non-local and non-equilibrium contributions to radiative transfer that make interpreting the measured diagnostics formidable \citep{Leenaarts_2020}. For the corona, the task is somewhat simpler and there is a wealth of suitable data from the likes of the Coronal Multi-Channel Polarimeter \citep{TOMetal2008} and the
Solar Dynamic Observatory \citep{PESetal2012}.  It is unclear why more attention hasn't been given to analysing the coronal Alfv\'enic waves present in the data. The measurements are challenging \citep[e.g.,][]{THUetal2014,WEBetal2018,tiwari_2021} but opportunities exist to compare the theoretical predictions discussed in preceding sections with observed wave behaviour. It is surely an area worthy of focus in the coming years. More attention may be focused on the observational behaviour of Alfv\'enic waves in the near future when DKIST \citep{Rimmele_2020,Rast_2021} and Solar Orbiter \citep{M_ller_2020} provide images of the chromosphere and corona at unprecedented spatial and temporal resolution. We wait with anticipation to see what wave physics these new facilities might uncover.

\acknowledgements{The authors are supported by a UKRI
Future Leader Fellowship (RiPSAW—MR/T019891/1). We also would like to thank
P. Antolin for patiently discussing a number of issues, and thank I. Arregui, M. Goossens, A. Hillier, R. Soler for valuable discussions. Further thanks are required to the two anonymous referees, whose insightful comments greatly improved the review. RJM would like to dedicate this review to the memory of Deenah Morton, without whose love and support this work and many others would not have been possible.
For the purpose of open access, the author(s) has applied a Creative Commons Attribution (CC BY) licence to any Author Accepted Manuscript version arising.

\medskip

\noindent Data and code used within this work is available at \url{https://github.com/Richardjmorton/FAL_models}.

\medskip

\noindent Figures have been created with the help of NumPy \citep{Numpy}, matplotlib \citep{Matplotlib} and IPython \citep{IPython}.}

\medskip
\noindent On behalf of all authors, the corresponding author states that there is no conflict of interest. 
\bibliographystyle{aa}


\end{document}